\documentclass[12pt,english]{article}
\usepackage[LGR,T1]{fontenc}
\usepackage[latin9]{inputenc}
\usepackage[a4paper]{geometry}
\usepackage{float}
\usepackage{textcomp}
\usepackage{amsmath}
\usepackage{graphicx}

\makeatletter

\DeclareRobustCommand{\greektext}{%
  \fontencoding{LGR}\selectfont\def\encodingdefault{LGR}}
\DeclareRobustCommand{\textgreek}[1]{\leavevmode{\greektext #1}}
\ProvideTextCommand{\~}{LGR}[1]{\char126#1}

\newcommand{\lyxmathsym}[1]{\ifmmode\begingroup\def\b@ld{bold}
  \text{\ifx\math@version\b@ld\bfseries\fi#1}\endgroup\else#1\fi}

\providecommand{\tabularnewline}{\\}

\usepackage{authblk}
\usepackage{upgreek}

\newcommand{\mykeywords}[1]{
	\small 
    \textbf{Keywords:} #1
}

\author[1,2]{Men Guo}
\author[2]{Gilad Orr}
\author[1,*]{Paul Ben Ishai}
\author[3]{Xia Zhao}
\author[2]{Shlomo Glasser}

\affil[1]{THz and Dielectrics Science Laboratory, Department of Physics, Ariel University, Ariel 407000,
Israel}
\affil[2]{Crystal Science Laboratory, Department of Physics, Ariel University, Ariel 407000, Israel}
\affil[3]{Department of High Voltage, China Electric Power Research Institute, Beijing 100192, People's
Republic of China}
\affil[*] {Corresponding author: Paul Ben Ishai, paulbi@arial.ac.il}
\date{}

\makeatother

\usepackage{babel}
\begin{document}
\title{Structural, optical, and dielectric properties of Cr-doped ZnO films
via DC magnetron sputtering}
\maketitle
\begin{abstract}
Cr-doped ZnO films were fabricated by a new but feasible method, that
is, annealing Cr-Zn layers deposited via DC magnetron sputtering in
air. Microstructures of the films were investigated using X-ray diffraction,
scanning electron microscopy, and atomic force microscopy, intrinsic
point defects were identified via photoluminescence spectroscopy,
and optical and dielectric properties were analyzed using a UV-vis
spectrophotometer and dielectric spectrometer, respectively. It was
found that the average grain sizes decrease (56.34\textminus 39.50
nm), the band gap increases (from 3.18 to 3.23 eV), and the transmittance
(at 600 nm) decreases (from 91\% to 83\%) with increasing Cr. Two
activation energies of conduction increase after doping Cr, indicating
enhanced temperature stability. At optimal Cr levels, ZnO films exhibit
high transmittance and conductivity, exhibiting potential for transparent
electrode development. This method can be extended to other doped
ZnO films, such as Al-doped ZnO transparent electrodes, to achieve
simultaneous improvements in transmittance, conductivity, and stability
for flexible and wearable applications.\\

\mykeywords{Transparent conductive oxides (TCO), Thin films, ZnO, Cr doping, Point defects}
\end{abstract}

\section{Introduction}

The transparent electrode is a crucial component of flexible and organic
optoelectronics. It has broad application prospects in wearable devices,
flat panel displays and energy sources \cite{Feng2024,Liu2021,Liu2021_2}.
Traditional indium tin oxide (ITO) electrode exhibits excellent optoelectronic
properties, with the sheet resistance ($R_{s}$) lower to $10\sim400\,\Omega/sq$
and the transmittance in the visible region higher than 80\% \cite{Kim1999,Lewis2000,Song2015}.
However, due to the limited availability of indium \cite{Kumar2010},
the high leakage current \cite{Hofmann201}, the brittleness \cite{Chen2001},
and environmental hazards, ITO cannot meet the requirements for future
scenarios, such as wearable devices and flexible displays \cite{Chen2022}.
It is still challenging to simultaneously improve the transmittance,
conductivity, and stability in environmental conditions with high
temperature, humidity, and mechanical stress.

ZnO is abundant, environmentally benign, and highly promising for
fabricating transparent electrodes \cite{Minami2013,Klingshirn2010,Lange2015}.
After doping Al, Ga, and/or In, the transmittance in the visible region
and the conductivity can exceed 90\% and $10^{3}\,S/cm$, respectively
\cite{Hsu2021,Das2021,Saxena2022}. Increasing the conductivity (to
$10^{4}\,S/cm$) while exhibiting the transmittance larger than 85\%
is difficult. On the other hand, when applied for flexible devices,
ZnO films face challenges including high temperature, high humidity,
high ultra-violet, sweat erosion, repeated bending and twisting. To
this end, passivation layers are applied on ZnO films to improve their
stability \cite{Pal2023,Islam2022}. However, it decreases the conductivity
and increases technical difficulty. 

Optical, electrical, and magnetic properties of ZnO-based materials
have been optimized by doping transition metals \cite{Wang2024,Ali2021,Jo2014}.
These have different valence states which assist in manipulating defects
in and the band structure of ZnO. To date, ZnO thin film doping has
focused on diluted magnetic properties and their mechanisms. Ueda
et al. observed room-temperature ferromagnetism (RTFM) in Co-doped
ZnO films \cite{Ueda2001}. Kittilstved et al. ascribed RTFM of Co-doped
ZnO films to the shallow donor zinc interstitials ($Zn_{i}$) \cite{Kittilstved2006}.
Sharma et al. and Ali et al. detected RTFM in Mn-doped ZnO thin films,
ascribed to $Mn^{2+}$ substituting for Zn at lattice sites and the
zinc vacancies ($V_{Zn}$) inducing ferromagnetic ordering, respectively
\cite{Ali2021,Sharma2003}. Nonetheless, optical and dielectric properties
are less investigated. 

Cr is known to resist severe environmental conditions and is a good
candidate for acting as a passivation layer when deposited on ZnO
films. In this study, we investigated Cr-doped ZnO films fabricated
through annealing double-layer Zn-Cr deposition in air. The effect
of Cr on the optical and dielectric properties of ZnO thin films were
systematically analysed from aspects of microstructure, band gap,
and point defects.

\section{Experimental procedures}

ZnO thin films were deposited on quartz substrates. The substrates
were wiped by a lint-free cloth soaked in soapy water to clean bulk
residues. After completely rinsing the substrates with distilled water,
they were cleaned twice in an ultrasonic bath with distilled water
for 35 minutes. Next, the substrates were rinsed thoroughly with ethanol
and cleaned twice in an ultrasonic bath with ethanol for 35 minutes.
Finally, the substrates were heated to 450°C for 30 min in air to
outgas. 

The deposition of metals was formed using a DC magnetron sputtering
machine (K575X Sputter Coater) at room temperature. The substrates
were positioned on a stage below the targets at a distance of 3 cm.
The base pressure was $7.00\times10^{-3}\,mbar$, the working current
was 50 mA, and the pressure was $9.00\times10^{-2}\,mbar$ when depositing.
At first, Cr (99.5\% purity) was deposited for \emph{x} seconds (\emph{x}
= 0, 30, 60, and 120), followed by depositing of Zn (99.99\% purity)
for 240 seconds. This resulted in a Cr-Zn double-layer structure on
the substrate. While sputtering, the substrate was rotated in order
to obtain a uniform deposition. Following the deposition, it was heated
at a rate of 10°C/min to 650°C for 60 min in air to be oxidized. Finally,
it was left at room temperature and allowed to cool naturally. The
samples were labelled CZO-0, CZO-30, CZO-60, and CZO-120, the number
referring to the sputtering time respectively. 

The phase composition and crystalline structure were characterized
using X-ray diffraction (XRD, Rigaku SmartLab SE) with Cu K\textgreek{a}
radiation at a step of 0.013°. The morphology was observed by scanning
electron microscopy (SEM, MAIA3 TESCAN) and the element distribution
was analyzed by energy dispersive spectrometry (EDS, Oxford Instruments).
The surface roughness and film thickness were measured using atomic
force microscopy (AFM, Bruker Dimension Icon). 

Raman scattering spectra were recorded in the range of $90\lyxmathsym{\textendash}1200\,cm^{-1}$
using an excitation wavelength of 532 nm. Photoluminescence (PL) spectra
were measured in the range of 325--600 nm using excitation wavelength
of 325 nm (Edinburgh FLS1000). The optical transmittance was measured
using a UV-vis spectrophotometer (Thermo Scientific). After depositing
Cr/Al electrodes, dielectric spectroscopy was measured using a dielectric
spectrometer (Novocontrol Technologies GmbH, Concept 80) with a signal
of 1 V (RMS) in a frequency range of $10^{-1}\lyxmathsym{\textendash}10^{6}\,Hz$
and a temperature range of $-100-150^{\circ}\text{C}$. The magnetization--magnetic
field ($M-H$) curves were measured using a commercial superconducting
quantum interference device (SQUID) magnetometer (Quantum Design MPMS
3).

\section{Results and discussion}

\subsection{Microstructural morphology}

\begin{figure}[H]
\begin{centering}
\includegraphics[height=6cm]{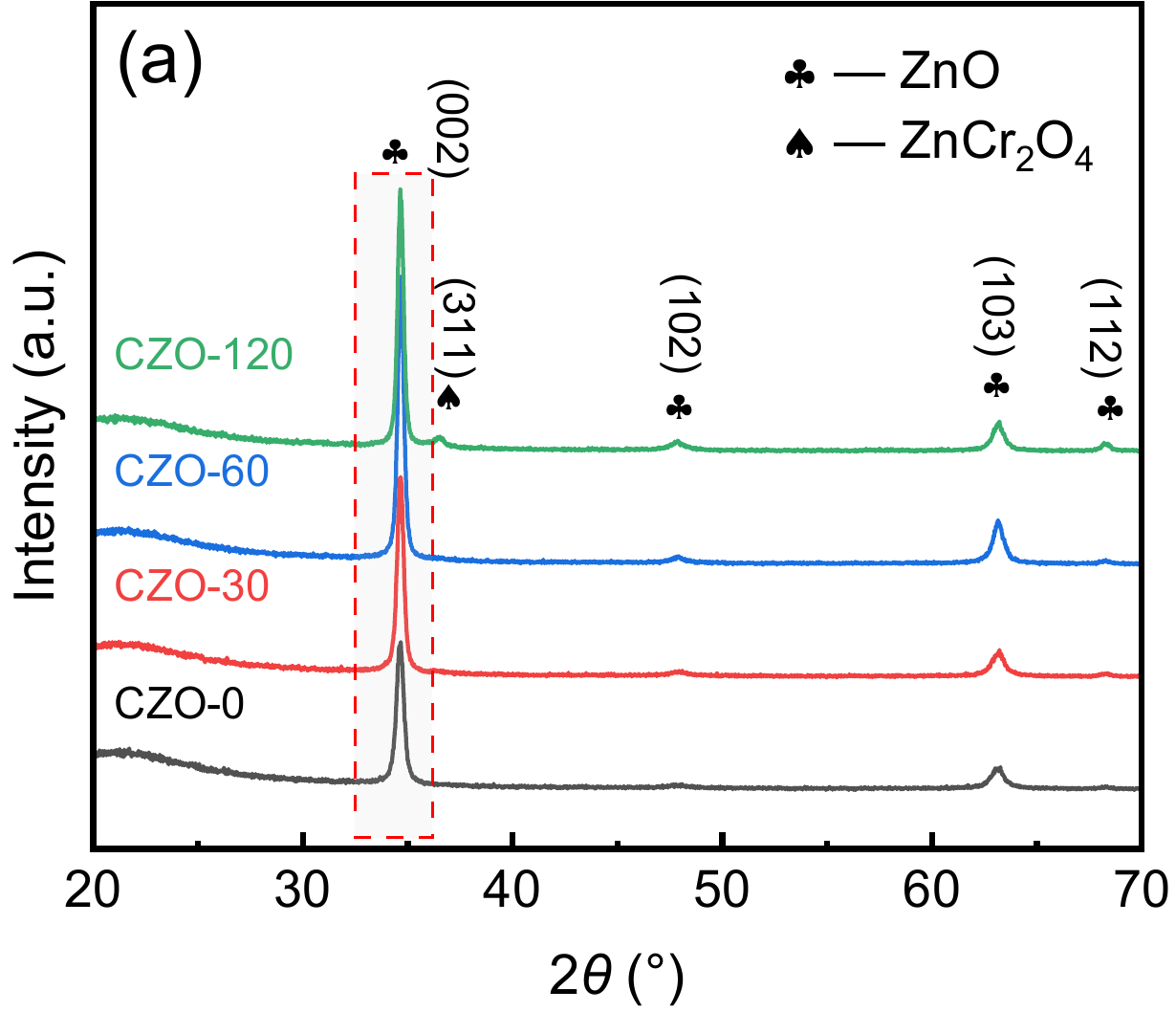}\includegraphics[height=6cm]{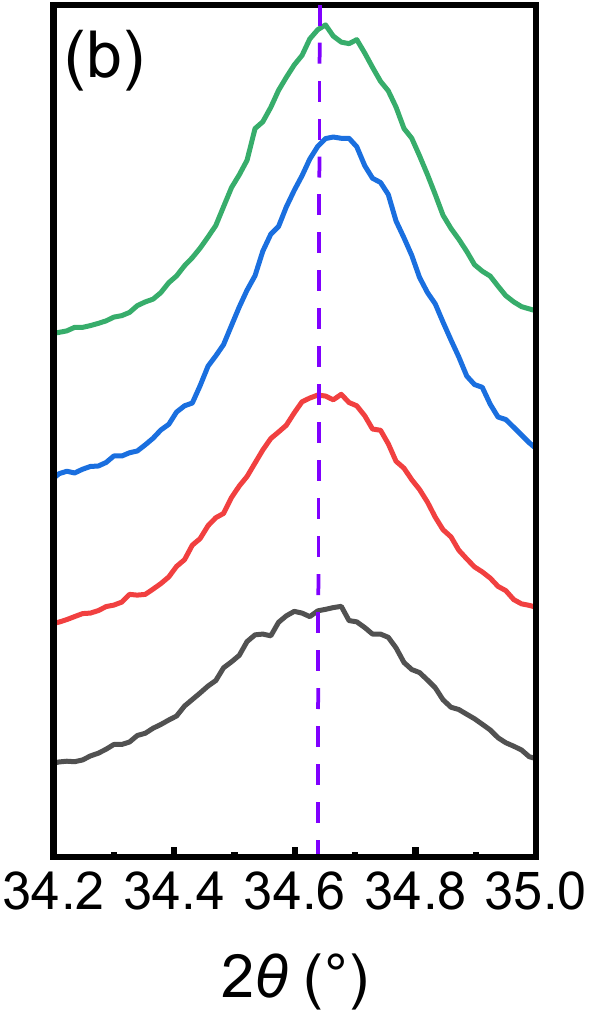}
\par\end{centering}
\caption{\label{fig:XRD-patterns-of_1}XRD patterns of Cr-doped ZnO films }

\end{figure}

XRD patterns of Cr-doped ZnO are shown in Figure \ref{fig:XRD-patterns-of_1}(a).
$2\theta$ is the diffraction angle. The strongest peak at $30-40{^\circ}$
corresponds to the hexagonal \emph{c}-oriented (002) plane of the
ZnO crystal (JCPDS Card No. 99-0111). After increasing the Cr content,
a secondary phase of $ZnCr_{2}O_{4}$ (JCPDS Card No. 73-1962) appears
at $36.54{^\circ}$ in CZO-120. The $ZnCr_{2}O_{4}$ phase in CZO-120
may contribute to reduced transmittance due to scattering. In the
zoomed-in view of (002) peaks, shown in Figure \ref{fig:XRD-patterns-of_1}(b),
$2\theta$ increases with increasing Cr content.

Fitting results of (002) plane in the XRD patterns are listed in Table
\ref{tab:Fitting-results-of-(002)}. According to Bragg\textquoteright s
law,
\begin{equation}
n\lambda=2d\sin\theta
\end{equation}

where \emph{n} is the diffraction order, $\lambda$ is the wavelength
of the X-ray, $d$ is the spacing of the crystal layers, and $\theta$
is the incident angle. The increase in $2\theta$ indicates that interplanar
spacings of the ZnO phases decrease, as shown in Table 1. The increase
in Cr content was verified by EDS measurement (See Figure S1 in Supplemental
information). The radius of $Zn^{2+}$ is $0.74\,\mathring{A}$, larger
than radii of common Cr ions ($0.615\,\mathring{{A}}$ of $Cr^{3+}$
and $0.44\,\mathring{A}$ of $Cr^{6+}$) \cite{shannon1976revised}.We
suggest that Cr ions dissolve into ZnO and substitute $Zn^{2+}$ ions,
resulting in the decrease in $d$ by shrinking cells. The Scherrer
equation \cite{scherrer1918abschatzungen} is $D=3K\lambda/(\text{FWHM}\cos\theta)$,
where $D$ is the crystallite size, $K$ is the dimensionless Scherrer
constant (\textasciitilde{} 0.9), $\lambda$ is the X-ray wavelength
($1.54056\,\mathring{A}$), FWHM is the full width at half maximum
of the peak (in radians, i.e., unitless angle when using the equation
\cite{orr2021crystalline}). $D$ increases from 22.4 nm to 28.7 nm
with increasing Cr content, suggesting that Cr facilitates the growth
of ZnO crystallites. These results are in accordance with previous
reports \cite{Iqbal2013,Al-Hardan2011,Hu2008}. 
\begin{table}[H]
\caption{\label{tab:Fitting-results-of-(002)}Fitting results of (002) plane
in XRD patterns of Cr-doped ZnO films.}

\centering{}%
\begin{tabular}{llccc}
\hline 
\noalign{\vskip0.1cm}
No. & $2\theta\,[\text{°}]$ & $d\,[\mathring{{A}}]$ & FWHM $[\text{°}]$ & D {[}nm{]}\tabularnewline
\hline 
\noalign{\vskip\doublerulesep}
CZO-0 & 34.599 & 2.5904 & 0.386 & 22.4\tabularnewline[\doublerulesep]
\noalign{\vskip\doublerulesep}
CZO-30 & 34.613 & 2.5894 & 0.342 & 25.5\tabularnewline[\doublerulesep]
\noalign{\vskip\doublerulesep}
CZO-60 & 34.651 & 2.5866 & 0.313 & 28.1\tabularnewline[\doublerulesep]
\noalign{\vskip\doublerulesep}
CZO-120 & 34.650 & 2.5866 & 0.307 & 28.7\tabularnewline[\doublerulesep]
\hline 
\end{tabular}
\end{table}

\begin{figure}[H]
\begin{centering}
\includegraphics[height=5cm]{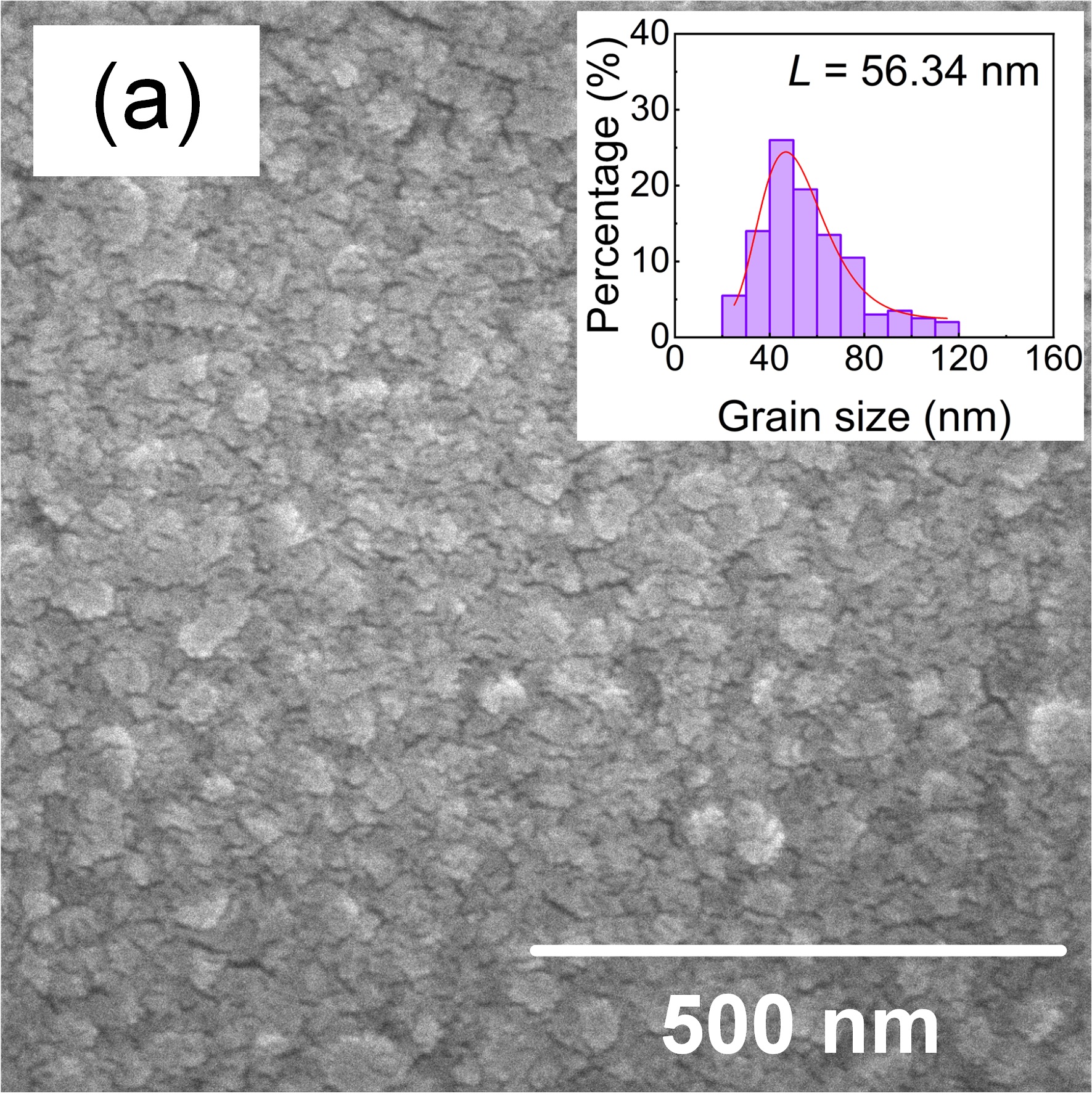} \includegraphics[height=5cm]{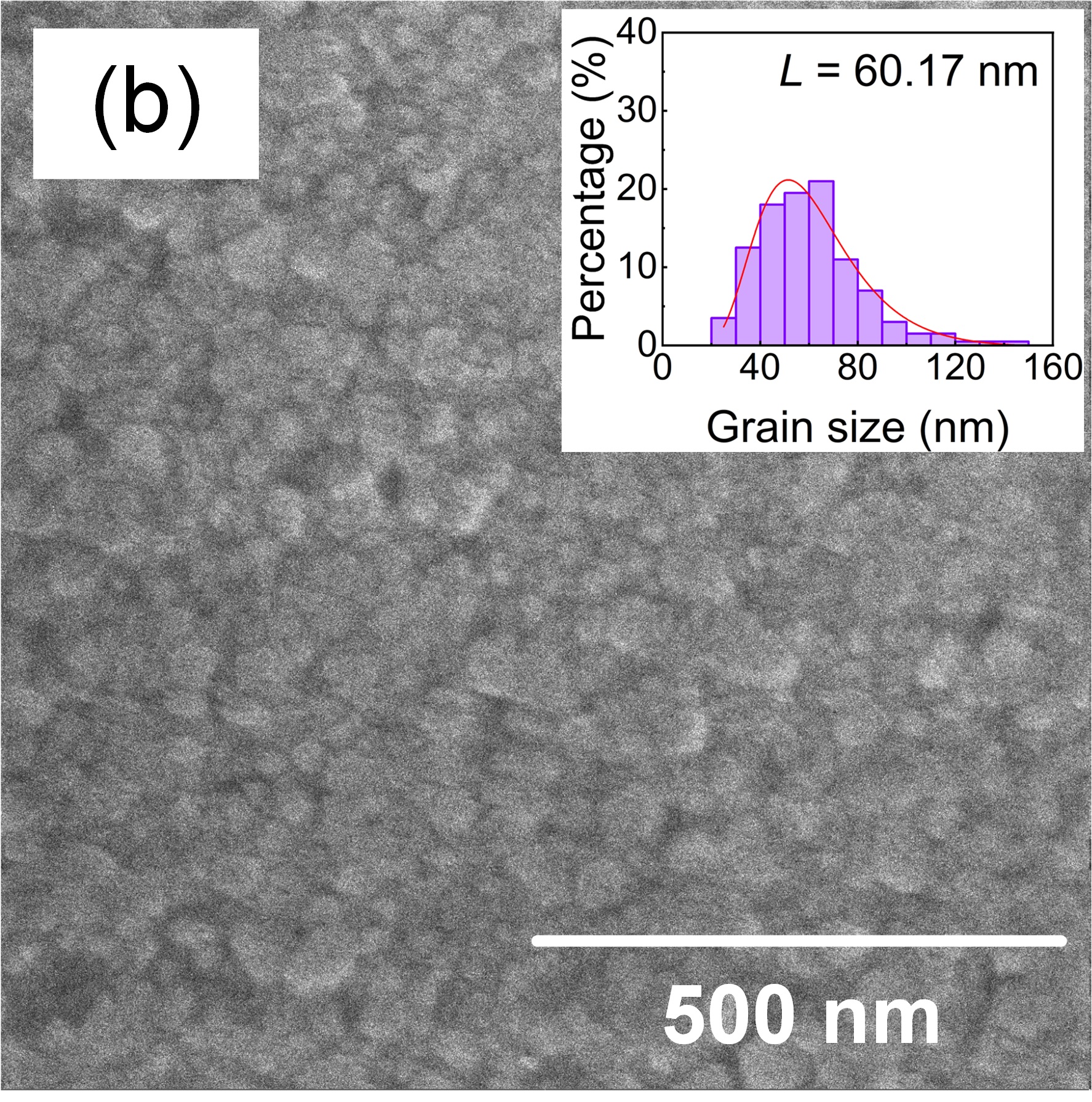}
\par\end{centering}
\begin{centering}
\includegraphics[height=5cm]{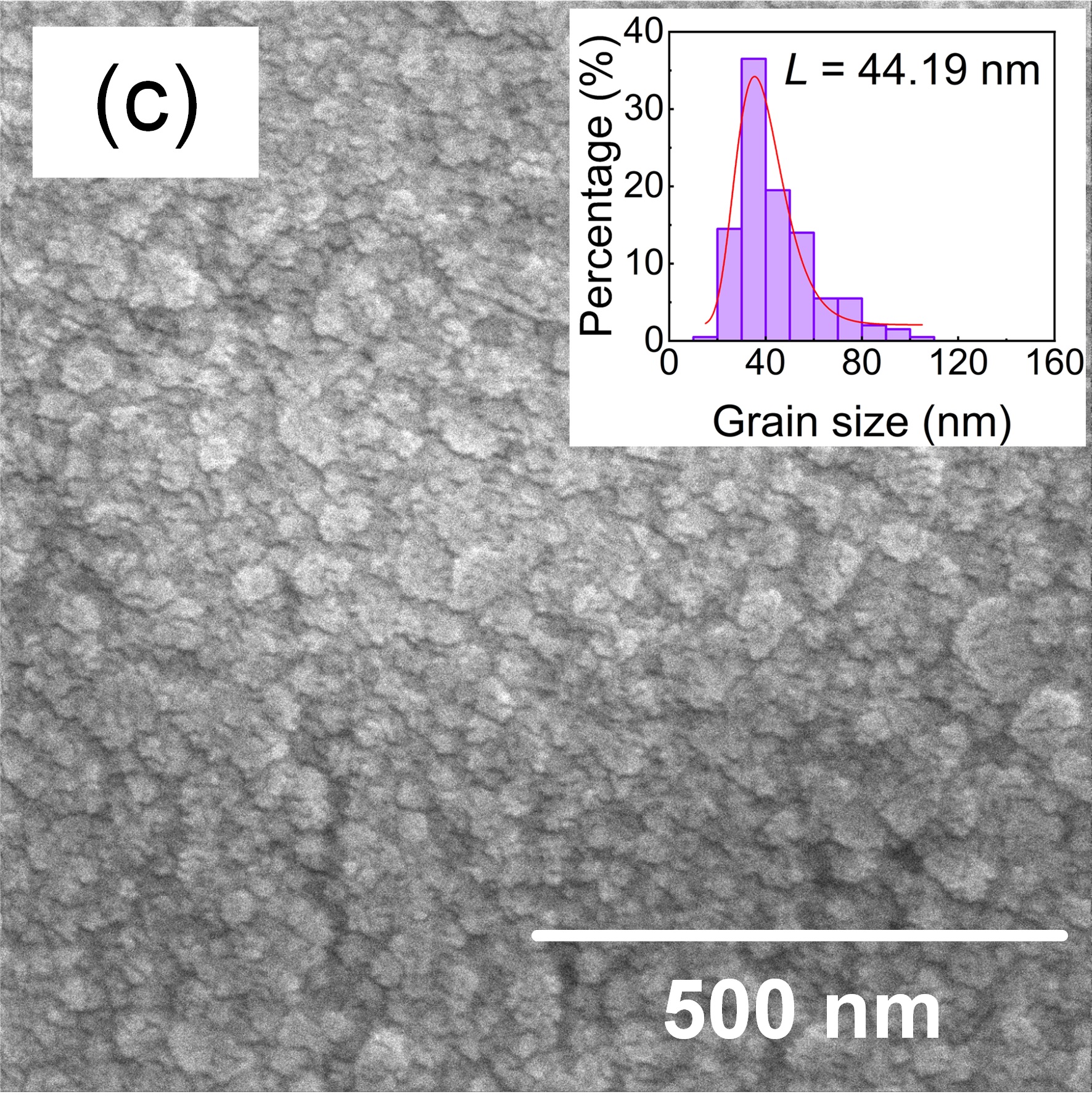} \includegraphics[height=5cm]{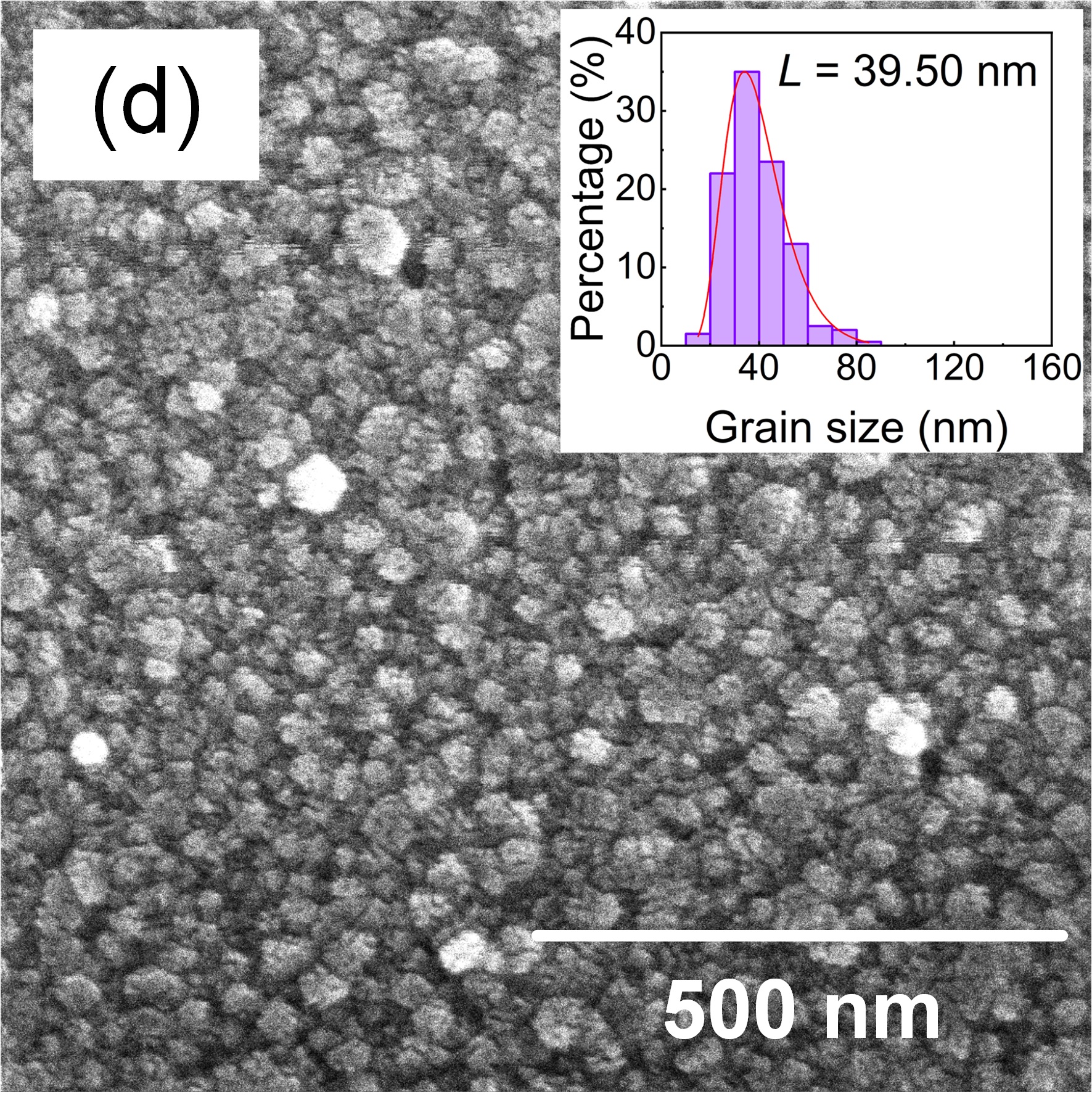}
\par\end{centering}
\caption{\label{fig:Top-view-SEM-images}Top-view SEM images of Cr-doped ZnO
films (a) CZO-0 (b) CZO-30 (c) CZO-60 (d) CZO-120.}

\end{figure}

SEM images of Cr-doped ZnO films are shown in Figure \ref{fig:Top-view-SEM-images}.
ZnO grains grow well in the samples and grain boundaries are distinguishable.
The grain size was measured by the linear intercept method \cite{Wurst1972}.
Insets of Figure \ref{fig:Top-view-SEM-images} are distributions
of grain sizes that follow the log-normal distribution. After doping
Cr, the average grain size (\emph{L}) increases from $56.34\pm19.4$
nm to $60.17\pm20.7$ nm. With increasing Cr content, \emph{L} monotonically
decreases to $39.50\pm11.8$ nm. On the other hand, the distribution
curves of CZO-60 and CZO-120 become narrower, suggesting that Cr acts
as a grain refiner at moderate levels, improving uniformity despite
smaller grain size. It is worth noting that the grain is not necessary
a crystallite and Cr has different effect on them in this case.

\begin{figure}[H]
\begin{centering}
\includegraphics[height=4cm]{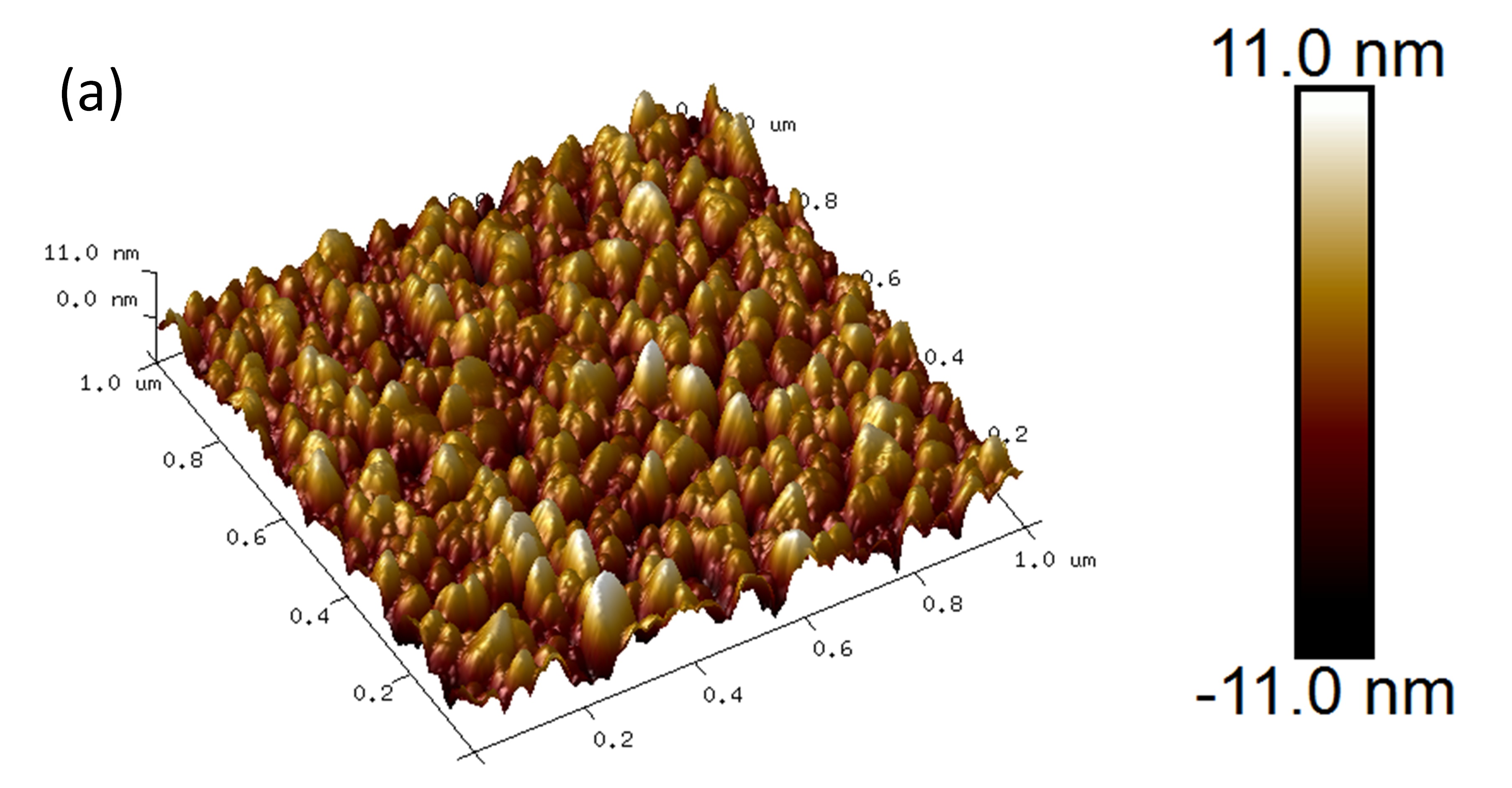} \includegraphics[height=4cm]{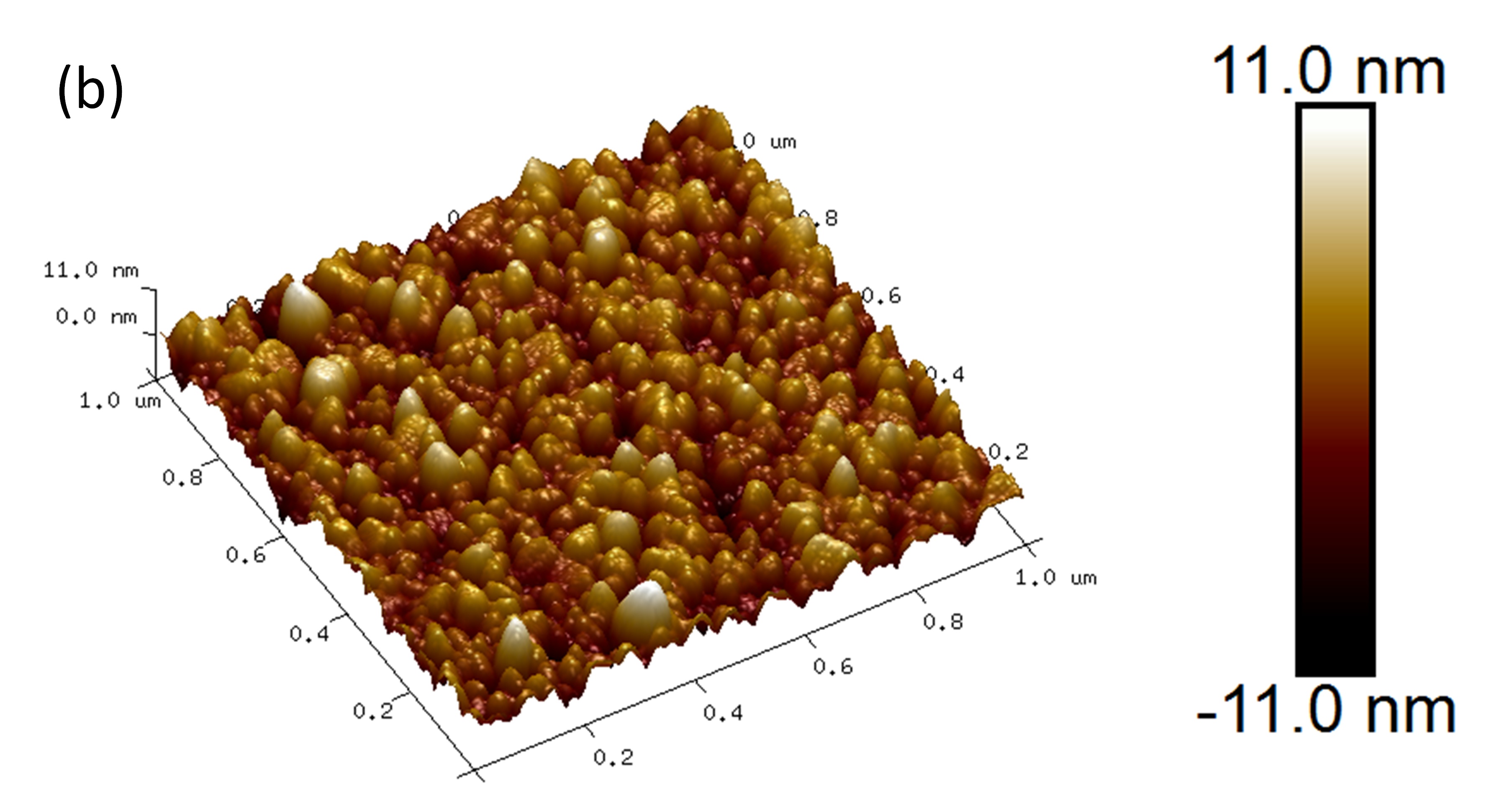}
\par\end{centering}
\begin{centering}
\includegraphics[height=4cm]{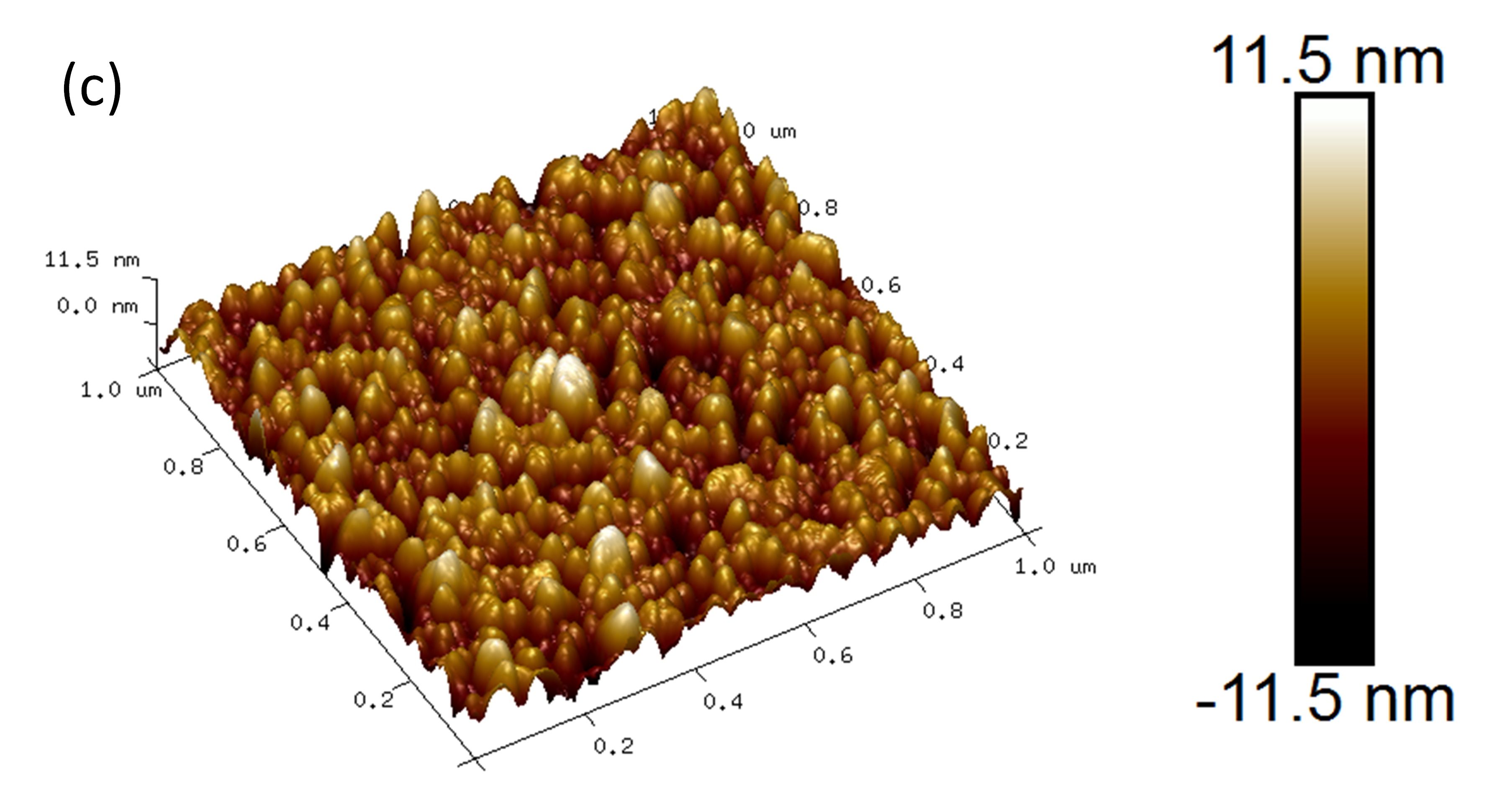} \includegraphics[height=4cm]{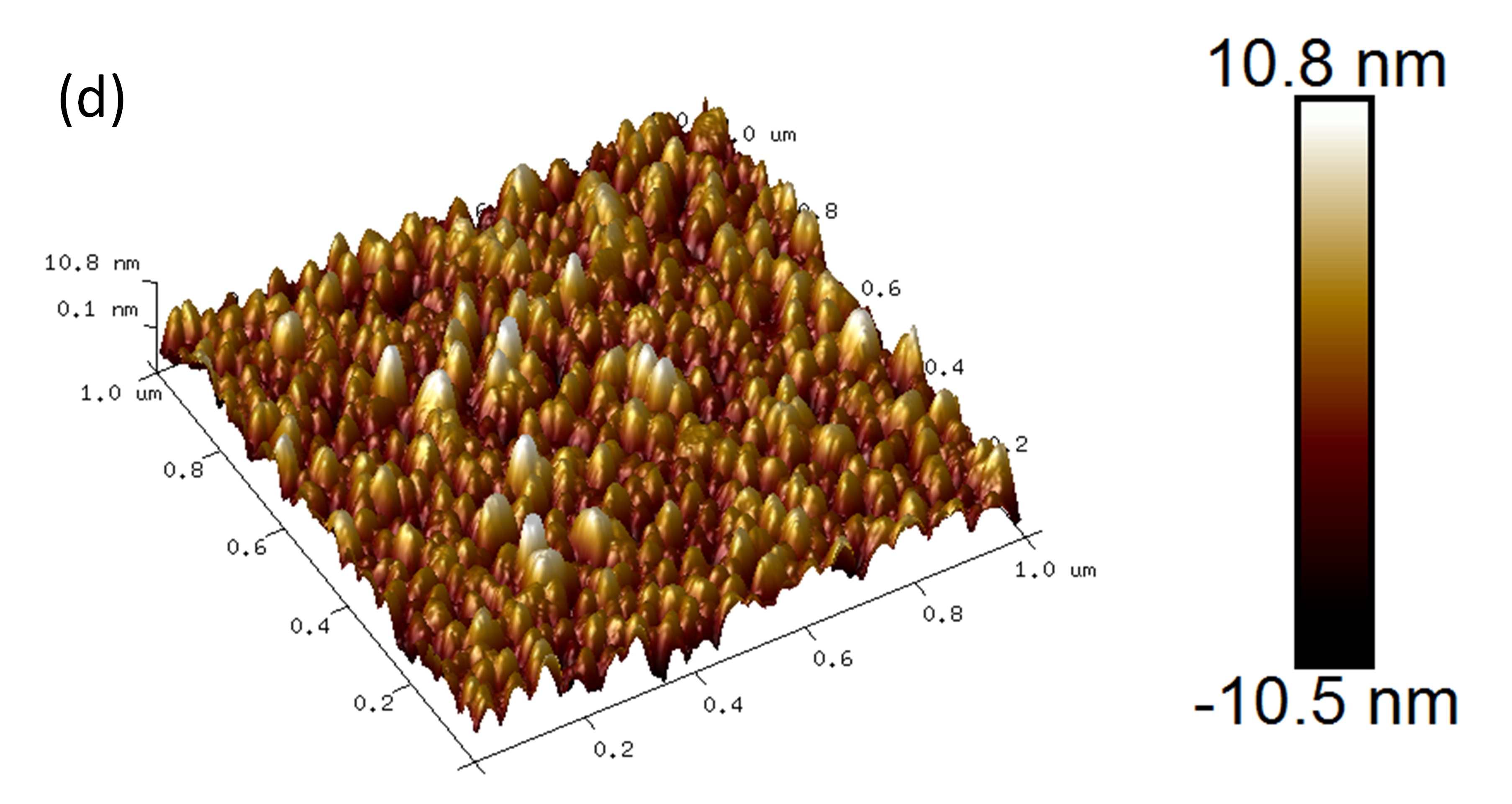}
\par\end{centering}
\caption{\label{fig:AFM-micrographs-of}AFM micrographs of Cr-doped ZnO films
(a) CZO-0 (b) CZO-30 (c) CZO-60 (d) CZO-120.}

\end{figure}

AFM micrographs with a scanning area of $1\,\mu m\times1\,\mu m$
for Cr-doped ZnO films are shown in Figure \ref{fig:AFM-micrographs-of}.
The undoped ZnO film (CZO-0) displayed a rugged surface with a root
mean square roughness ($R_{q}$) of 2.99 nm. For ZnO films with small
amount of Cr doping (CZO-30), $R_{q}$ decreases to 2.58 nm. With
increasing Cr doping, $R_{q}$ increases to 3.04 nm and 2.93 nm for
CZO-60 and CZO-120, respectively. The thickness (\emph{t}) was also
measured using AFM (shown in Figure S2). $t$ is 53.8, 63.9, 41.3,
and 25.3 nm for CZO-0, CZO-30, CZO-60, and CZO-120, respectively.

\subsection{Analysis of point defects}

\begin{figure}[H]
\begin{centering}
\includegraphics[height=5cm]{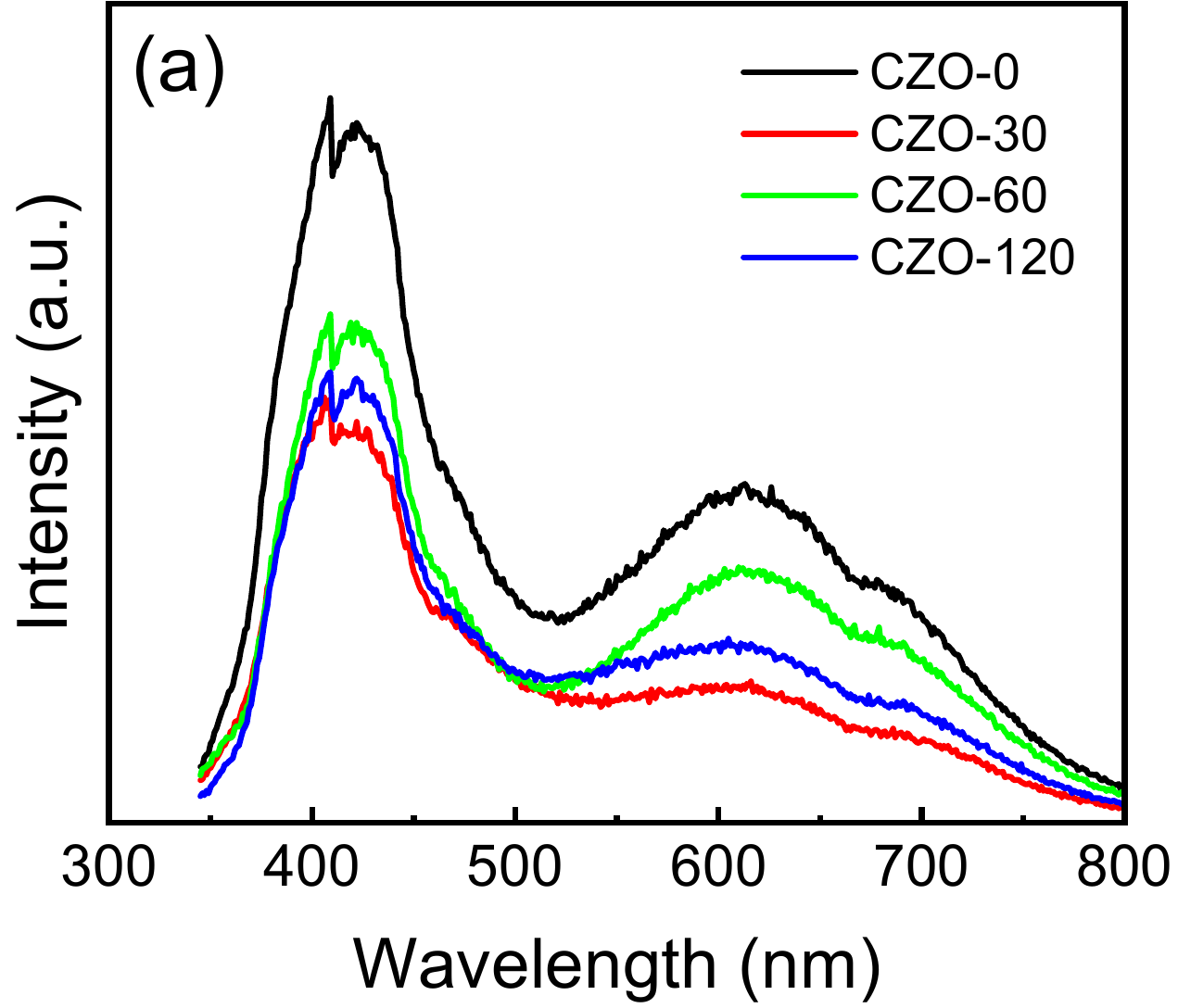} \includegraphics[height=5cm]{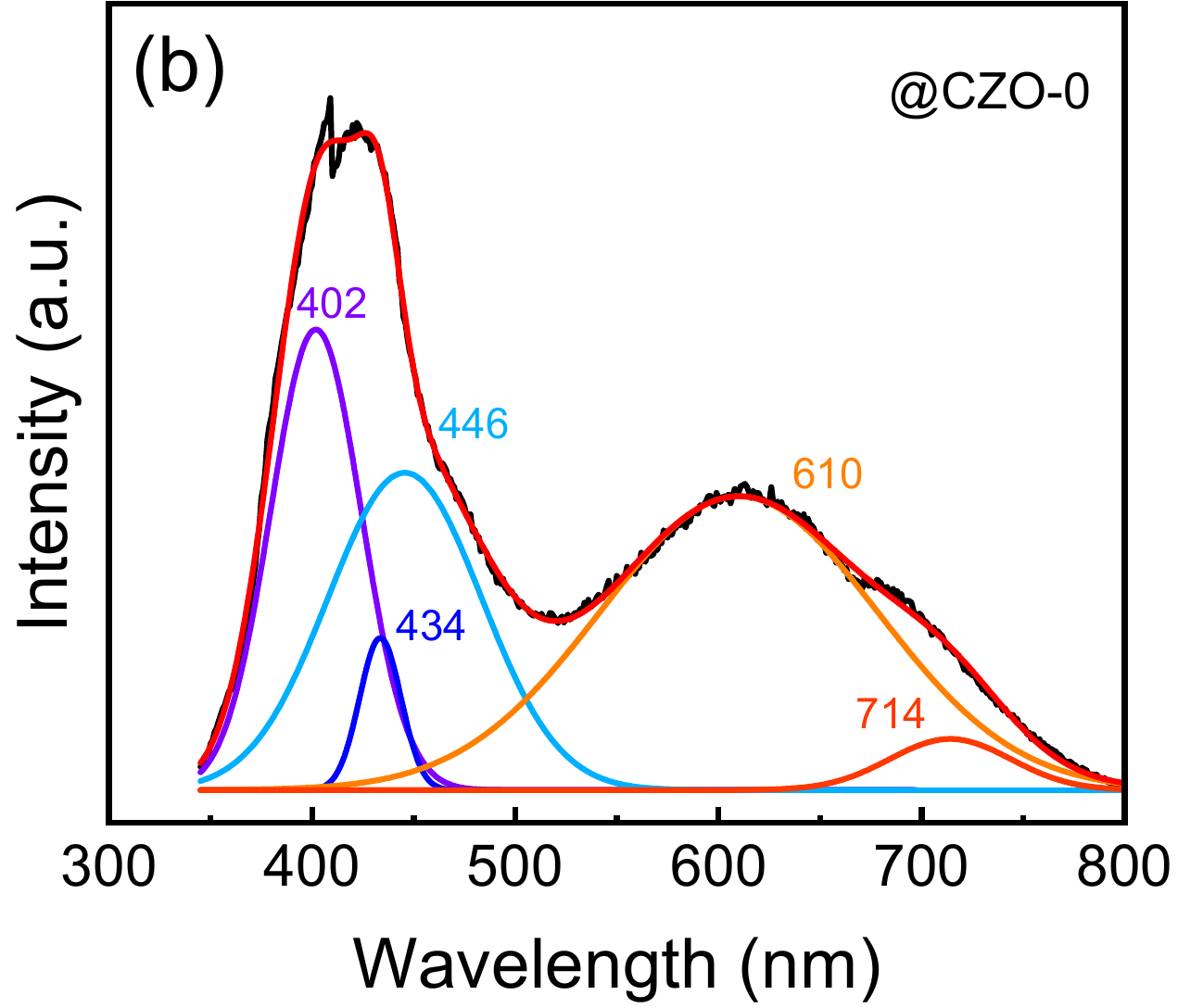}
\par\end{centering}
\begin{centering}
\includegraphics[height=5cm]{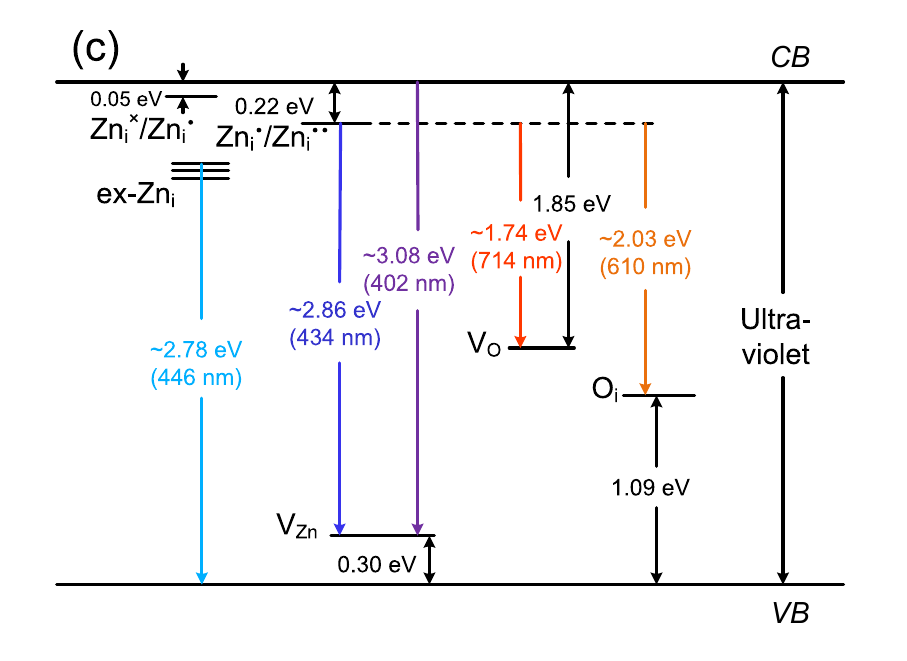} \includegraphics[height=5cm]{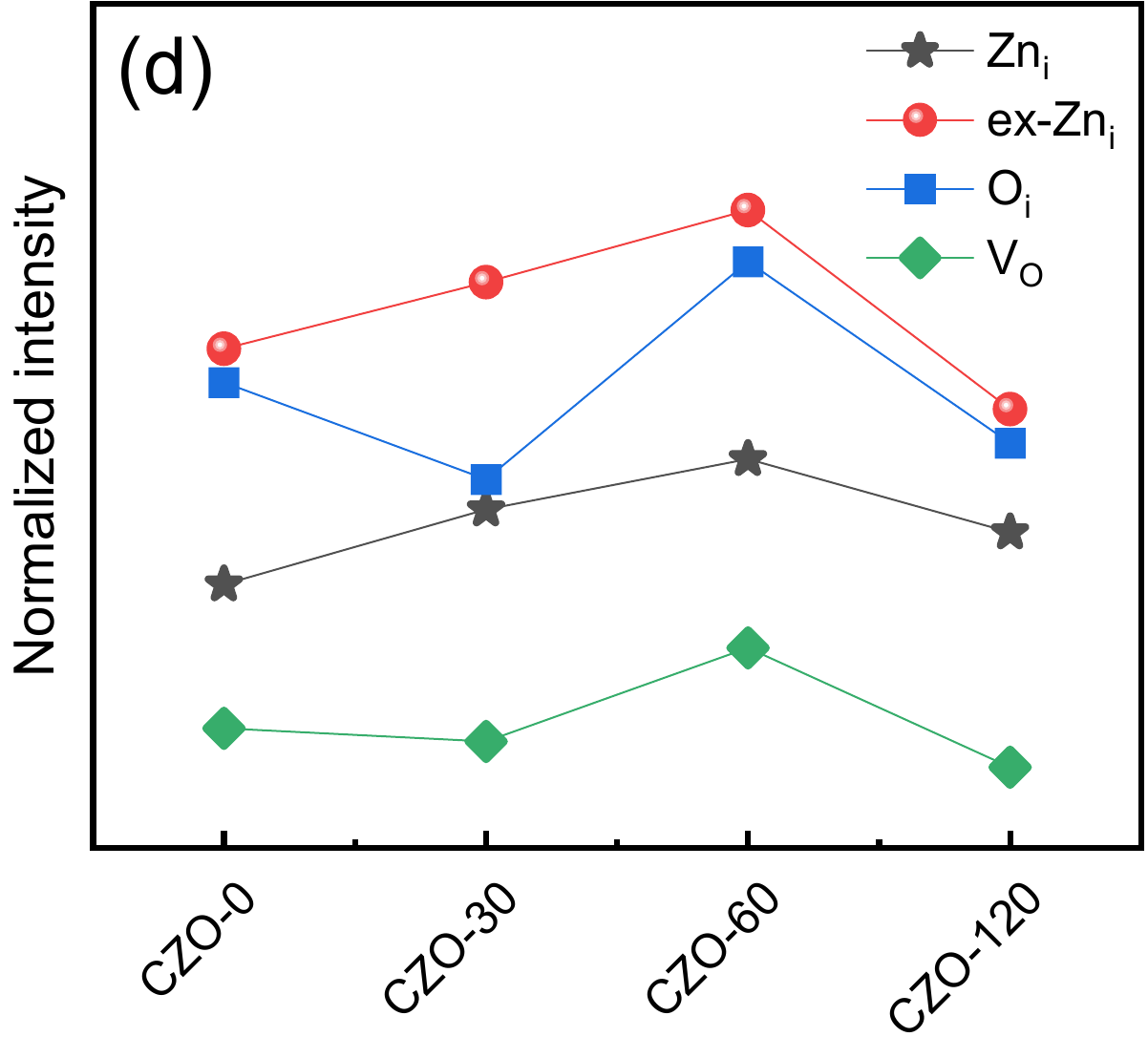}
\par\end{centering}
\caption{\label{fig:Photoluminescence-spectra-of}Photoluminescence spectra
of Cr-doped ZnO films measured at room temperature.}

\end{figure}

PL spectra of the Cr-doped ZnO films are shown in Figure \ref{fig:Photoluminescence-spectra-of}(a).
After doping Cr, the intensities of PL spectra show a decreasing tendency.
The deconvolution was performed for the spectra and five peaks were
detected using Gaussian fitting, as shown in Figure \ref{fig:Photoluminescence-spectra-of}(b).
The violet emission peak at $402\pm0.5$ nm (3.08 eV) is well consistent
with the energy interval from the bottom of the conduction band to
$V_{Zn}$ level (3.06 eV) \cite{Lin2001}. The blue emission at 434
nm (2.86 eV) corresponds to the transition from $Zn_{i}$ to $V_{Zn}$
level \cite{Ahn2009}. Another blue emission at $445.6\pm3.4$ nm
(2.78 eV) was attributed to electron transitions from extended $Zn_{i}$
states (complex defects or localized $Zn_{i}$ states) to the valence
band \cite{Zeng2010}. The orange emission at $610.0\pm0.9$ nm (2.03
eV) is relevant to transitions from $Zn_{i}$ to the oxygen interstitial
($O_{i}$) \cite{Ahn2009,Li2004}. The red emission at $714.1\pm1.5$
nm (1.74 eV) is close to the interval from $V_{O}$ to $Zn_{i}$ or
$V_{O}$ to the conduction band \cite{Kumar2014,Kumar2013}. Figure
\ref{fig:Photoluminescence-spectra-of}(c) illustrates energy levels
of detected defects, showing transitions responsible for PL emissions.
Based on the intensity of the 402-nm peak ($V_{Zn}$), PL spectra
were normalized. Changes in other defects with Cr content are shown
in Figure \ref{fig:Photoluminescence-spectra-of}(d). With increasing
Cr content, $Zn_{i}$ increases in the films up to CZO-60 due to Cr
substitution. Cr-O bond is much stronger than Zn-O. When Cr is oxidized,
it can snatch oxygen from ZnO and facilitate the production of $Zn_{i}$
and $V_{O}$, as expressed in Equations \ref{eq:2} and \ref{eq:3}.
$Cr^{3+}$ substitutes $Zn^{2+}$ at lattice sites through Equation
\ref{eq:4}, which decreases interplanar spacings in ZnO (observed
by XRD). However, $Zn_{i}$ decreases in CZO-120 due to $ZnCr_{2}O_{4}$
formation. At a certain amount of Cr (CZO-60), these defects reach
the highest level, but $O_{i}$ and $V_{O}$ do not increase monotonically.
After further doping Cr, all defects drop to low levels (CZO-120).
Because the films are ultrathin, no Raman signals from the films were
detected (shown in Figure S5).
\begin{equation}
ZnO+V_{i}^{\times}\rightarrow Zn_{i}^{\times}+V_{O}^{\times}+0.5O_{2}\label{eq:2}
\end{equation}

\begin{equation}
2Cr+1.5O_{2}\xrightarrow{\Delta}Cr_{2}O_{3}\label{eq:3}
\end{equation}

\begin{equation}
Cr_{2}O_{3}\xrightarrow{2ZnO}2Cr_{Zn}^{\bullet}+2O_{O}^{\times}+2e'+0.5O_{2}\label{eq:4}
\end{equation}

\subsection{Optical properties}

\begin{figure}[H]
\begin{centering}
\includegraphics[height=5cm]{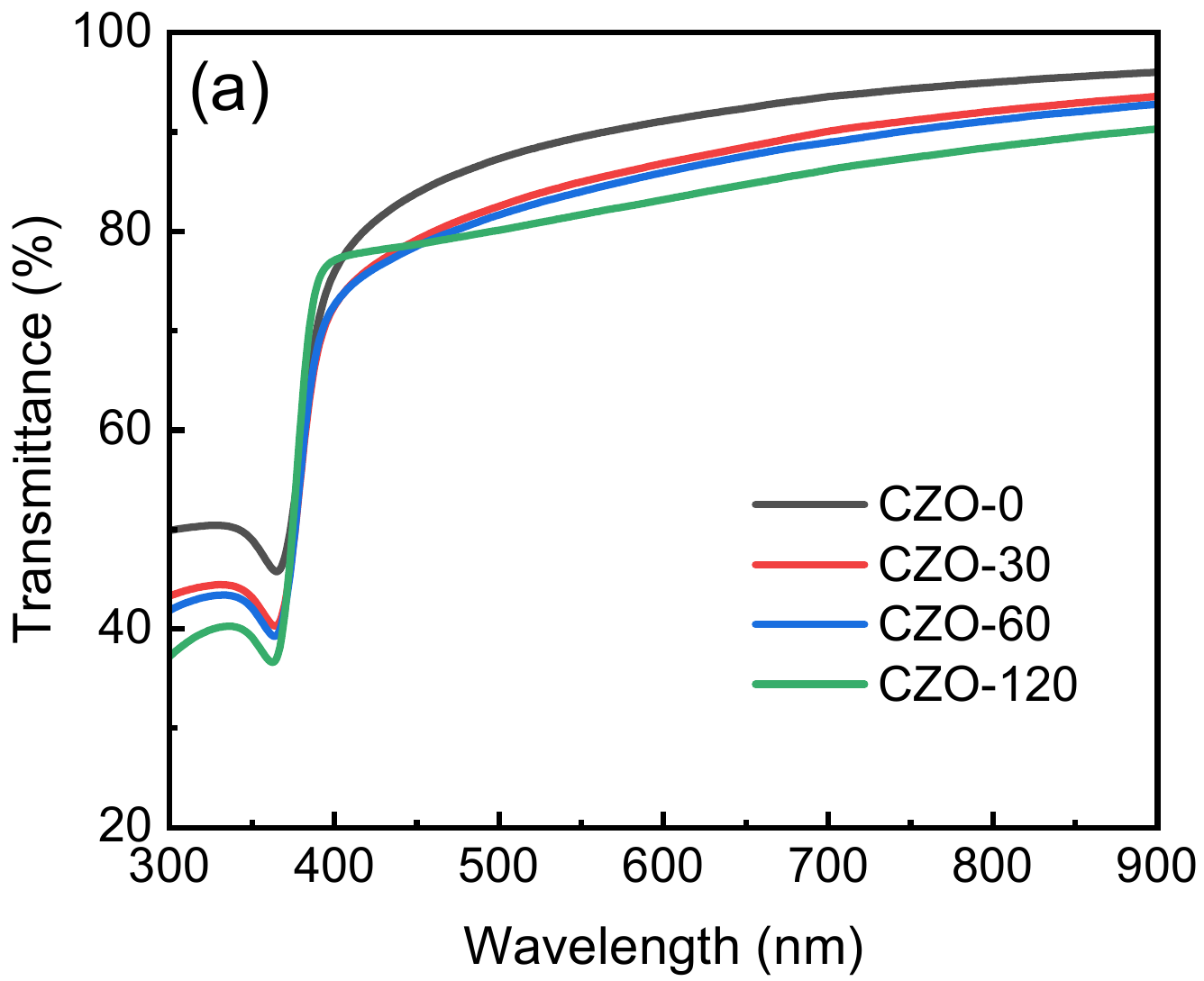} \includegraphics[height=5cm]{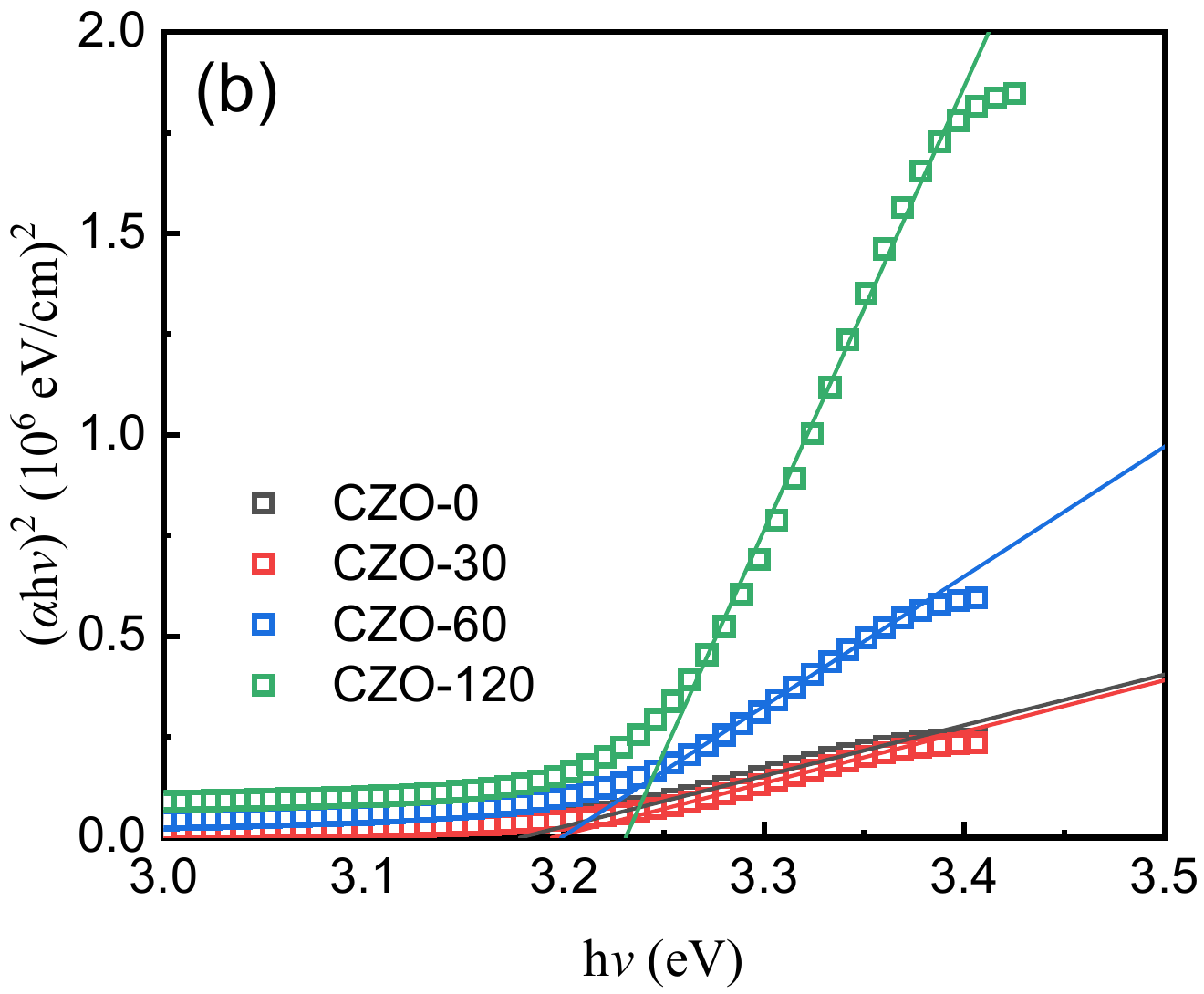}
\par\end{centering}
\begin{centering}
\includegraphics[height=5cm]{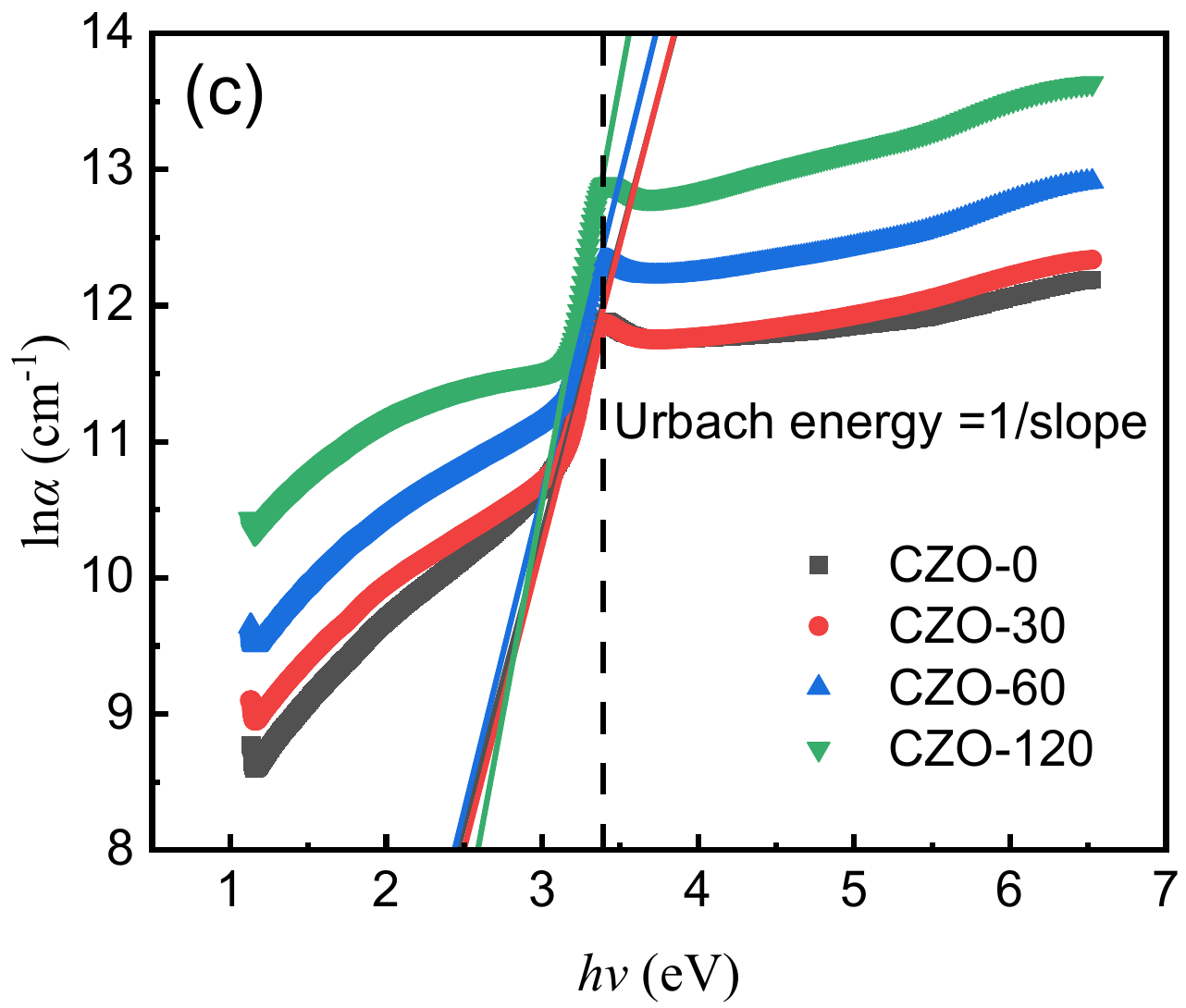} \includegraphics[height=5cm]{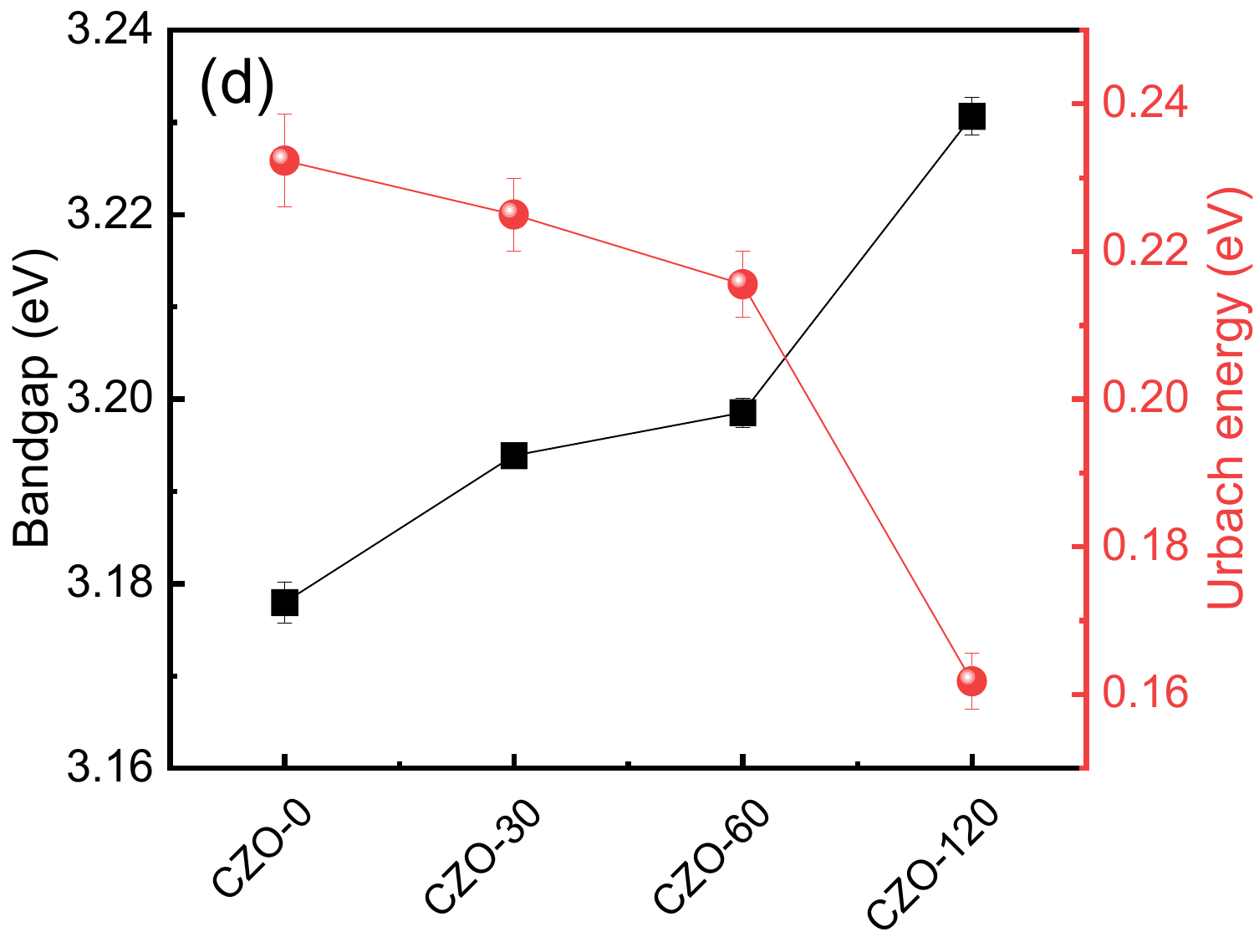}
\par\end{centering}
\caption{\label{fig:(a)-Transmission-spectra}(a) Transmission spectra and
(b) Tauc's plot, (c) logarithm of absorption coefficients as a function
of photon energy, and (d) band gap and Urbach energy of Cr-doped ZnO
films.}

\end{figure}

Transmission spectra of Cr-doped ZnO thin films are shown in Figure
\ref{fig:(a)-Transmission-spectra}. After doping Cr in ZnO films,
the transmittance ($T$) in the visible light region (380\textminus 750
nm) decreases from 91\% to 87\% (600 nm), shown in Figure \ref{fig:(a)-Transmission-spectra}(a).
With increasing Cr content, $T$ further decreases to 83\% (600 nm).
$T$ dramatically decreases in the UV region due to the strong fundamental
absorption. The relationship between the absorption coefficient ($\alpha$)
and the incident photon energy ($h\nu$) reads \cite{Pankove1975}:

\begin{equation}
\alpha h\nu=A(h\nu-E_{g})^{m}\label{eq:5}
\end{equation}

\begin{equation}
\alpha=-\frac{1}{t}\ln\frac{T}{(1-R)^{2}}\label{eq:6}
\end{equation}
 where $A$ is a constant, $E_{g}$ is the bandgap, $m=1/2,2$ for
direct and indirect allowed transitions, respectively, $t$ is the
film thickness, and $R$ is the reflectance (ignored in this study).
$E_{g}$ can be obtained through the Tauc\textquoteright s plot \cite{Tauc1972},
i.e., extrapolating the straight-line portion of $h\nu-(\alpha h\nu)^{2}$
plot to the energy axis, as shown in Figure \ref{fig:(a)-Transmission-spectra}(b),
assuming that the energy gap is direct. With increasing Cr content
in ZnO films, $E_{g}$ increases from 3.18 eV to 3.23 eV, as listed
in Table \ref{tab:Band-gap-and}. Furthermore, due to the very small
$t(<65\,nm)$ and finite $\alpha$, $T$ above $E_{g}$ is non-zero
(Equation \ref{eq:6}, $T\approx exp(-\alpha t)$).

In semiconductors and insulators, the fundamental absorption edge
below the energy band gap, i.e., the Urbach tail, increases exponentially
\cite{urbach1953long}. The Urbach tail does not follow the sharp
Tauc edge, but shows an exponential slope \cite{rai2013analysis}:
\begin{equation}
\alpha(h\nu)=\alpha_{0}\exp\left(\frac{h\nu-E_{0}}{E_{U}}\right)
\end{equation}
 where $E_{0}$ and $\alpha_{0}$ are the characteristic parameters
of the materials, and $E_{U}$ is the Urbach energy. The $\ln(\alpha)$
vs. $h\nu$ plot is shown in Figure \ref{fig:(a)-Transmission-spectra}(c)
and $E_{U}$ can be obtained through the slope of the straight line.
$E_{U}$ is 0.23, 0.22, 0.22, and 0.16 eV in the ZnO films with increasing
Cr content.

Due to the sequential deposition of Cr-Zn layers followed by annealing,
a gradual diffusion of Cr into the Zn layer is expected, resulting
in a Cr concentration gradient across the film thickness (gradient
doping). If this is the case, then in the bottom region, the content
of Cr is higher, which is beneficial to higher conductivity (Equation
\ref{eq:4}). As extending out, the content of Cr would gradually
decrease, which helps maintain higher transmittance by inhibiting
the carrier scattering or absorption. Therefore, this doping gradient
would be expected to modify optical and dielectric properties by providing
a passivation-like effect near the surface while preserving the conductive
nature of the bulk ZnO. If the gradient doping is formed in the Cr-ZnO
films, the tendency of $E_{g}$ would be different from previous studies
\cite{Iqbal2013,Hu2008}. The changes of $E_{U}$ are in accordance
with the tendency of $E_{g}$, as shown in Figure \ref{fig:(a)-Transmission-spectra}(d).

\subsection{Dielectric properties}

The impedance ($Z^{*}$) of ZnO films can be expressed as

\begin{equation}
Z^{*}=1/j\omega C^{*}=1/j\omega C_{0}\varepsilon^{*}
\end{equation}

where $j$ is the square root of \textminus 1, $\omega$ is the angular
frequency, $C_{0}$ is the geometric capacitance, and $C^{*}=C'-jC''$
and $\varepsilon^{*}=\varepsilon'-j\varepsilon''$ are complex capacitance
and permittivity, respectively. The electrodes for dielectric measurements
were deposited on the same surface of ZnO films. Because it is difficult
to find the geometry of thin films, instead of permittivity, capacitances
are present for dielectric analysis. 

\begin{figure}[H]
\begin{centering}
\includegraphics[height=5cm]{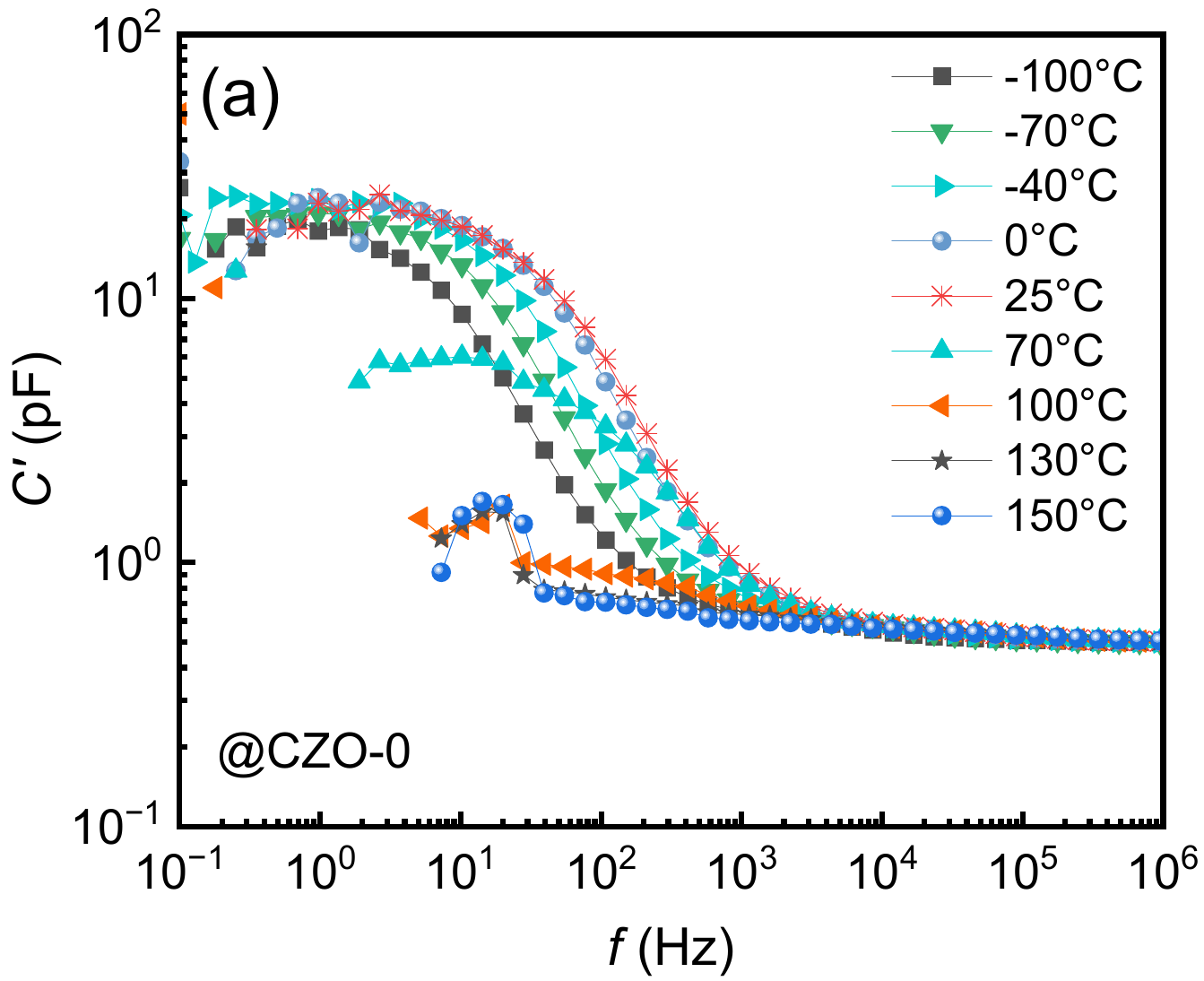} \includegraphics[height=5cm]{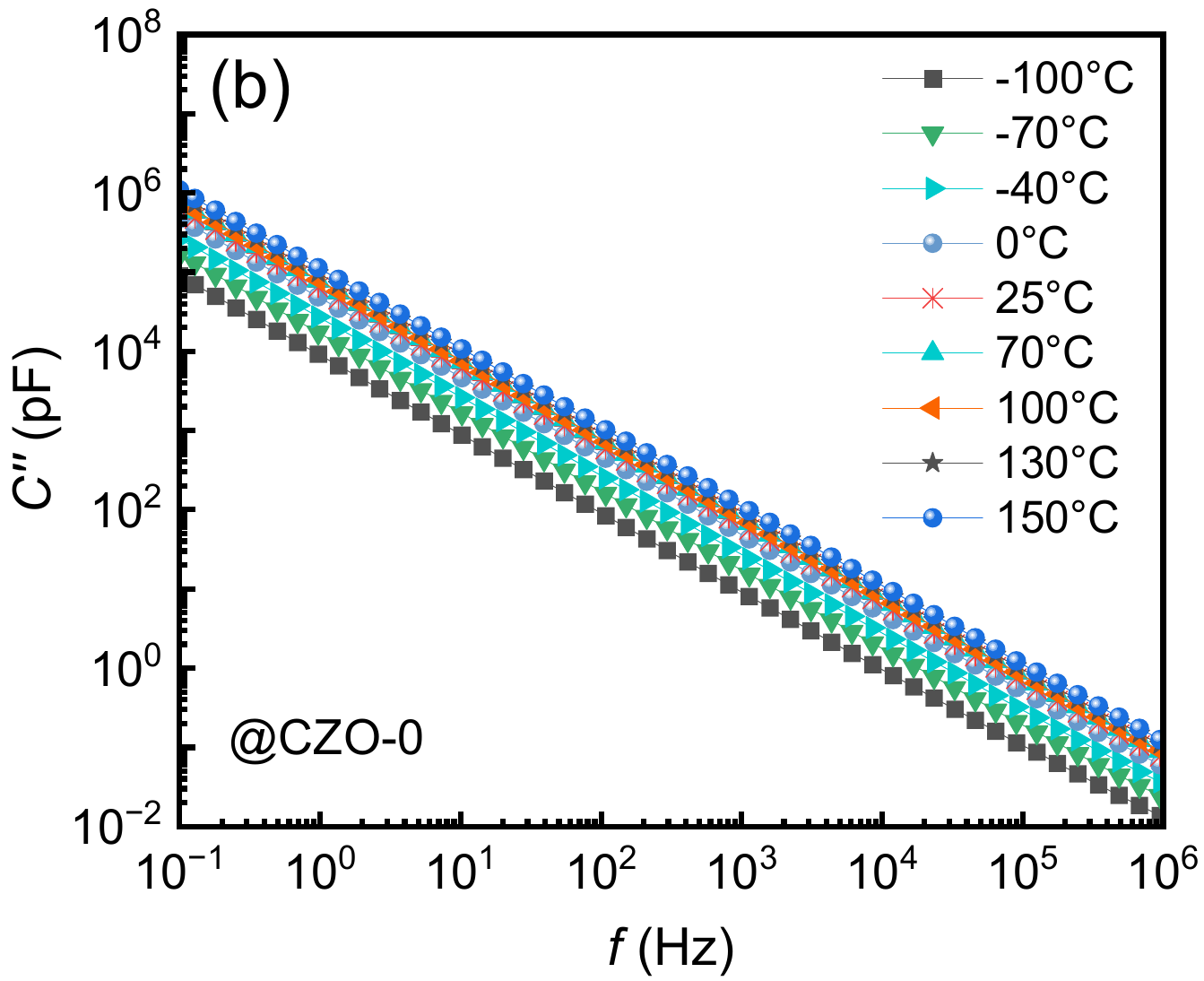}
\par\end{centering}
\begin{centering}
\includegraphics[height=5cm]{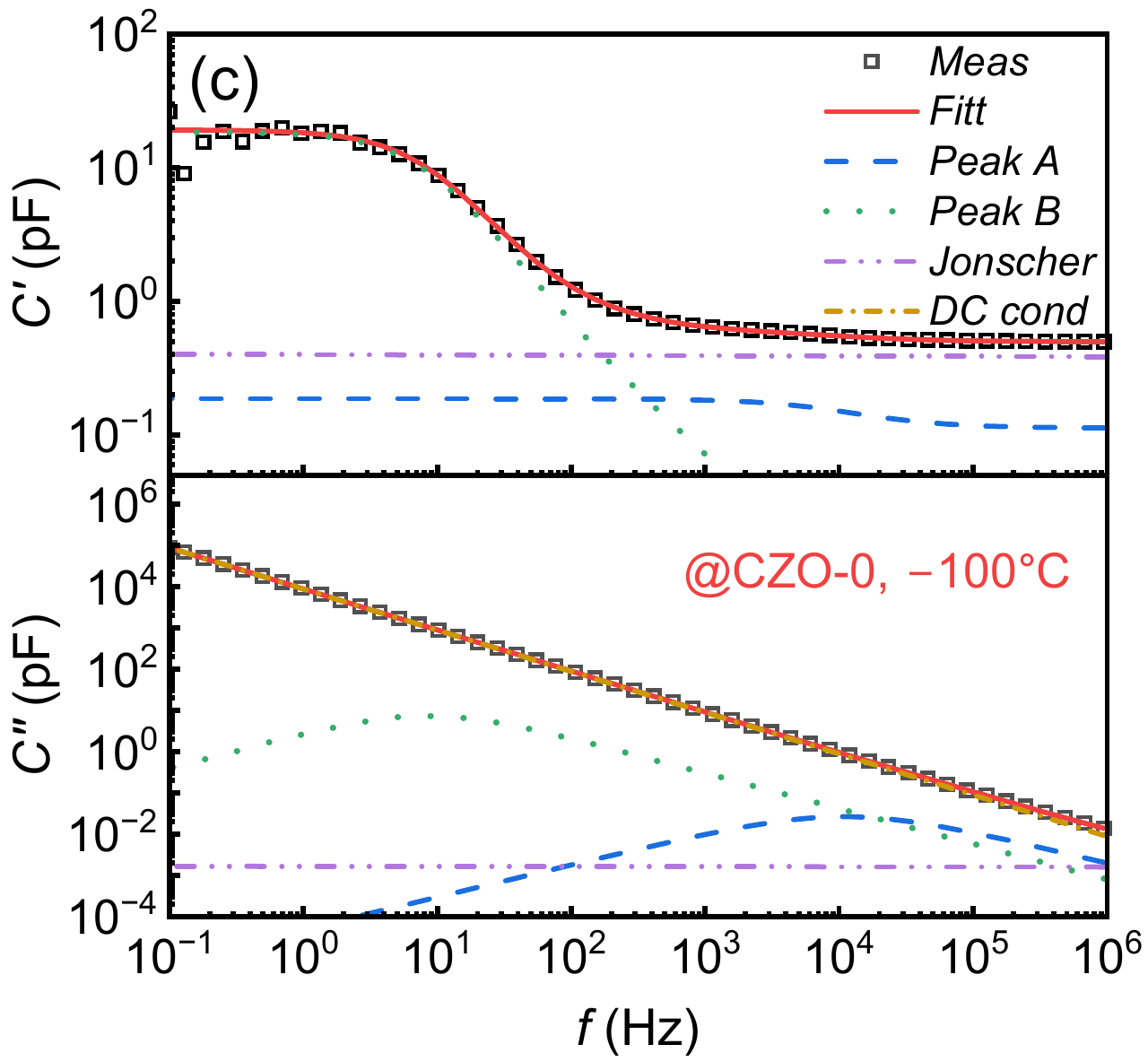} \includegraphics[height=5cm]{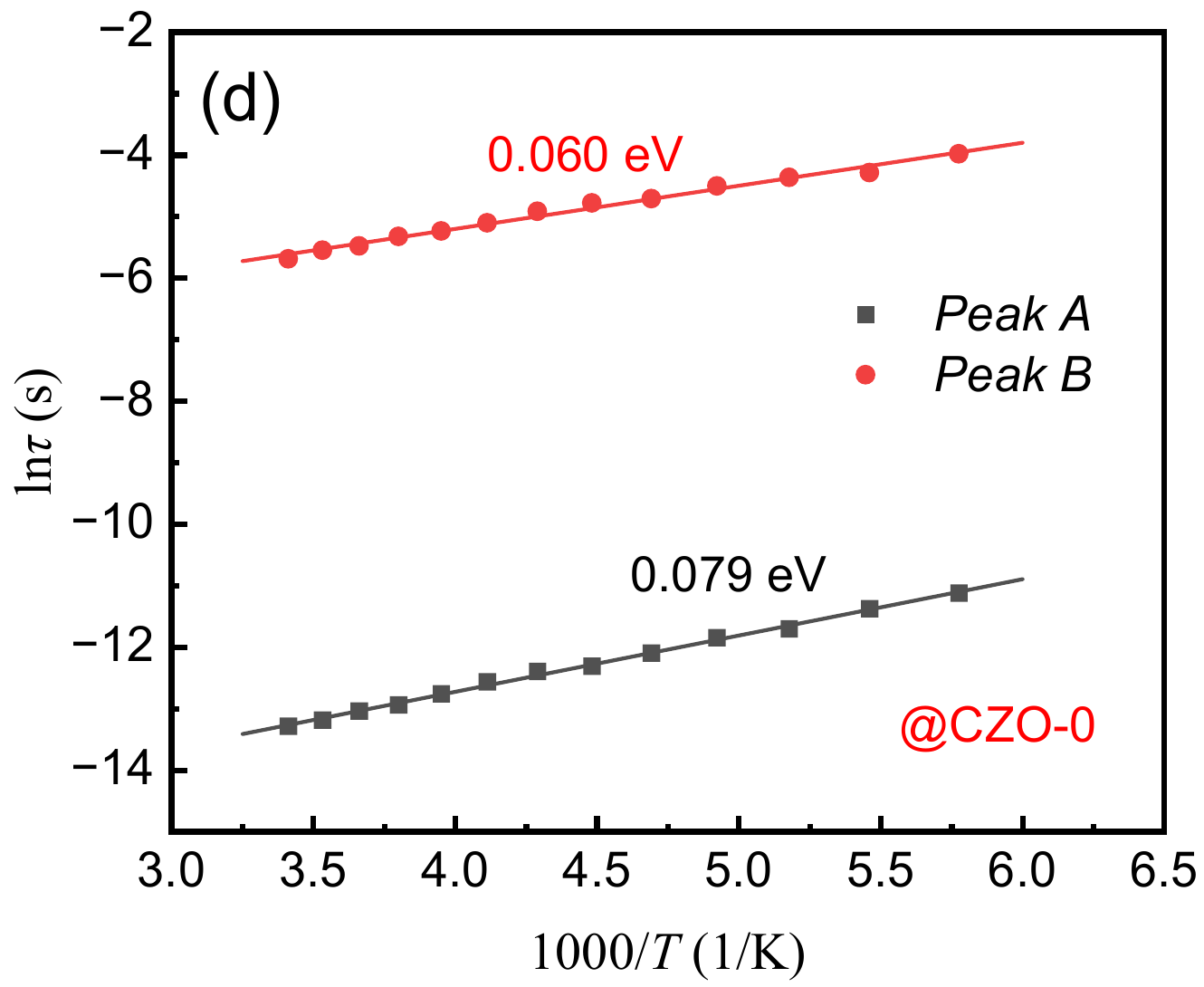}
\par\end{centering}
\caption{\label{fig:Dielectric-spectra-of}Dielectric spectra of CZO-0 film
(a) the real part $C'$ (b) the imaginary part $C''$ and (c) fitting
results at \textminus 100°C (d) Arrhenius behaviours of relaxations.}

\end{figure}

Dielectric spectra of CZO-0 are shown in Figure \ref{fig:Dielectric-spectra-of}.
To verify the influence of quartz, electrodes with the same configuration
were deposited on a quartz substrate, for which dielectric properties
were measured (Figure S3). Compared with ZnO films, the capacitance
of the quartz substrate can be ignored. In Figure \ref{fig:Dielectric-spectra-of}(a),
some stages exhibit in the spectra, indicating some relaxations. In
Figure \ref{fig:Dielectric-spectra-of}(b), only parallel lines are
present in the spectra, instead of loss peaks, suggesting dominant
DC conduction. With increasing temperature, the relaxations shift
in the high-frequency direction, similar to the Debye relaxation. 

The Cole-Cole model was employed to fit measurement results of dielectric
spectra \cite{Cole1942}:
\begin{equation}
C^{*}=C_{0}\left(\varepsilon_{\infty}+\frac{\sigma_{dc}}{j\omega\varepsilon_{0}}+\underset{i}{\sum}\frac{\Delta\varepsilon_{i}}{1+(j\omega\uptau_{i})^{\alpha_{i}}}\right)
\end{equation}

where $\varepsilon_{\infty}$ is the high-frequency permittivity,
$\sigma_{dc}$ is the $DC$ conductivity, $\Delta\varepsilon_{i}$
is the strength, $\uptau_{i}$ is the relaxation time, and $\alpha_{i}\,(\in(0,1))$
is the broadening parameter of the $i^{\text{th}}$ relaxation. In
the low-frequency region, for relaxations situated at frequencies
lower than the measurement window the Jonscher power law was adopted
for fitting power law behaviours \cite{Jonscher1996}, that is, $\varepsilon^{\ast}=K(j\omega)^{-n}$
, where $K$ and $n$ are temperature-dependent constants and $n\in(0,1)$.
Fitting results of CZO-0 at $-100{^\circ}C$ are shown in Figure \ref{fig:Dielectric-spectra-of}(c).
Two relaxations, named \emph{Peak A} and \emph{Peak B} are detected.
As shown in Figure \ref{fig:Dielectric-spectra-of}(d), the results
of the fitting reveal well defined Arrhenius behaviour for the characteristic
relaxation times of the processes, i.e.,
\begin{equation}
\uptau_{i}=\uptau_{0i}\exp\left(\frac{E_{i}}{kT}\right),\,i=1,2,3,...
\end{equation}

where $\uptau_{i}$ is the mean relaxation time of the \emph{i}\textsuperscript{th}
relaxation peak, $\uptau_{0i}$ is its precursor, and $E_{i}$ is
the activation energy for the relaxation \cite{BenIshai2013,Orr2022}.
Activation energies of relaxations in the films are listed in Table
\ref{tab:Band-gap-and}. In CZO-0, $E_{A}$ and $E_{B}$ are close
to the defect level of $Zn_{i}^{\times}/Zn_{i}^{\bullet}$, i.e.,
the first ionization of $Zn_{i}$. $E_{B}$ of CZO-30 and $E_{A}$
of CZO-120 approximate to the level of $Zn_{i}^{\bullet}/Zn_{i}^{\bullet\bullet}$,
i.e., the second ionization of $Zn_{i}$ \cite{Guo2025}. The activation
energy in 0.15\textminus 0.17 eV may be associated with extended $Zn_{i}$
states.

\begin{figure}[H]
\begin{centering}
\includegraphics[height=6cm]{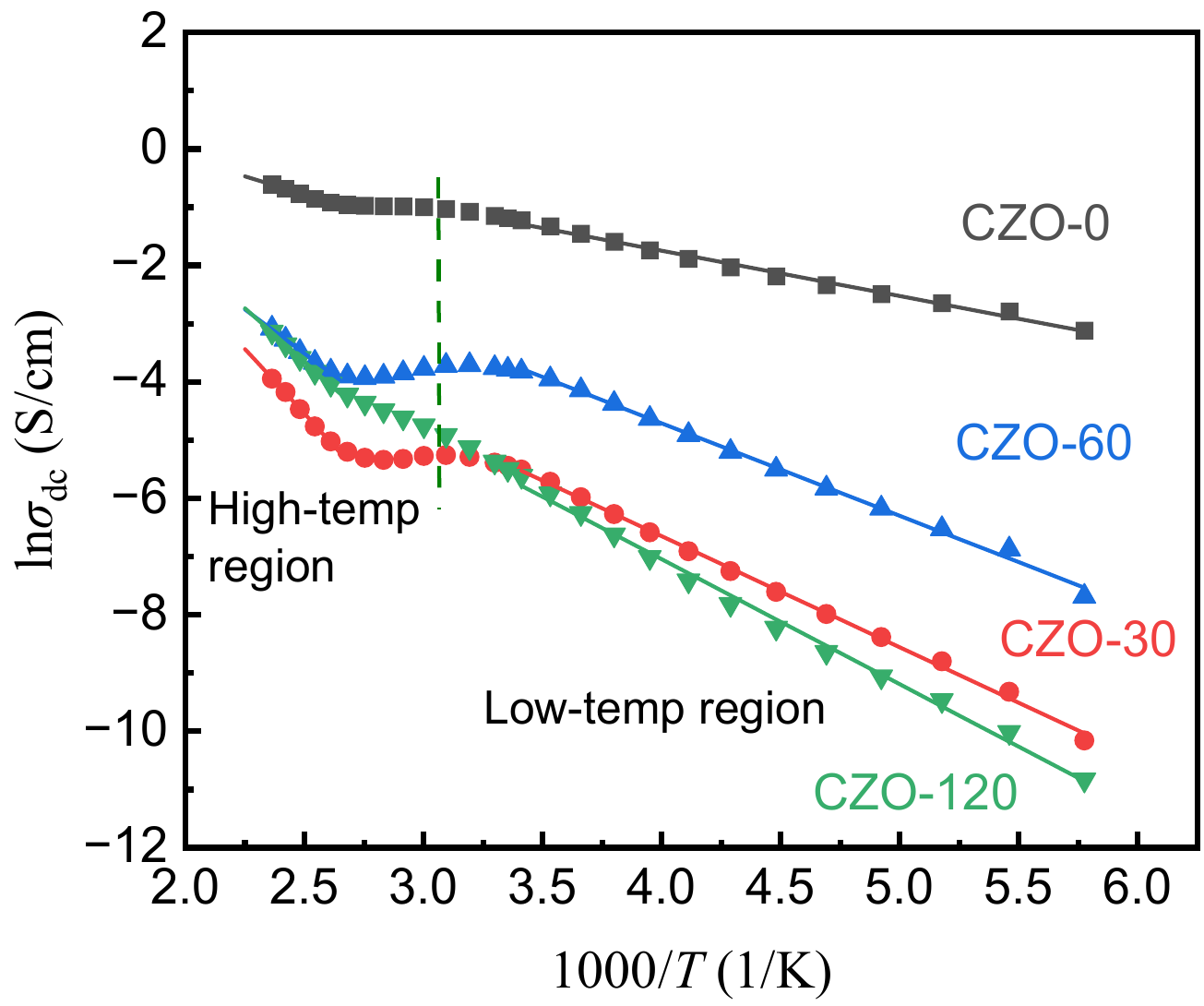}
\par\end{centering}
\caption{\label{fig:Arrhenius-behaviours-of}Arrhenius behaviours of the DC
conduction in Cr-doped ZnO films.}

\end{figure}

The $DC$ conductivity $\sigma_{dc}$ vs. the temperature is shown
in Figure \ref{fig:Arrhenius-behaviours-of}. $\sigma_{dc}$ satisfies:
$\sigma_{dc}=\sigma_{0}exp(-E_{dc}\lyxmathsym{\textfractionsolidus}kT)$,
where $E_{dc}$ is the activation energy of the $DC$ conduction \cite{Orr2022,orr2018high,Jonscher1999}.
There are two distinct activation processes respectively at low temperatures
(< 40°C) and high temperatures (> 90°C). Activation energies of the
two processes ($E_{LT}$ and $E_{HT}$) are also listed in Table \ref{tab:Band-gap-and}.
After doping Cr, the activation energies in both regions become much
larger. Given comparable thicknesses of the films and the same electrode
configuration, we can assume that $C_{0}$ is similar for all the
films. Therefore, changes of $\sigma_{dc}$ with Cr doping content
are also reflected in Figure \ref{fig:Arrhenius-behaviours-of}. After
doping Cr and at a higher content of Cr, $\sigma_{dc}$ decreases
(CZO-30 and CZO-120). But for an optimal content of Cr, $\sigma_{dc}$
is at a higher level (CZO-60).

\begin{table}[H]

\caption{\label{tab:Band-gap-and}Band gap and activation energies of Cr-doped
ZnO films (eV)}

\begin{centering}
\begin{tabular}{cccccc}
\hline 
No. & $E_{g}$ & $E_{A}$ & $E_{B}$ & $E_{LT}$ & $E_{HT}$\tabularnewline
\hline 
CZO-0 & $3.18\pm0.002$ & $0.079\pm0.001$ & $0.060\pm0.002$ & $0.067\pm0.002$ & $0.11\pm0.004$\tabularnewline
CZO-30 & $3.19\pm0.001$ & $0.17\pm0.002$ & $0.22\pm0.001$ & $0.16\pm0.002$ & $0.38\pm0.009$\tabularnewline
CZO-60 & $3.20\pm0.002$ & $0.16\pm0.001$ & $0.15\pm0.002$ & $0.14\pm0.002$ & $0.26\pm0.013$\tabularnewline
CZO-120 & $3.23\pm0.002$ & $0.20\pm0.002$ & $0.17\pm0.001$ & $0.18\pm0.004$ & $0.32\pm0.008$\tabularnewline
\hline 
\end{tabular}
\par\end{centering}
\end{table}

After doping Cr, $E_{g}$ of ZnO films increases while the transmittance
in the visible region decreases. According to the analysis of PL spectra,
shallow donors, i.e., $Zn_{i}$ and ex-$Zn_{i}$, increase in CZO-30
and CZO-60 and $V_{O}$ also increases in CZO-60. The reason why the
transmittance decreases is the scattering of photons by these donors.
With further increasing Cr doping, $L$ decreases to 39.50 nm. In
other words, there are more grain boundaries in CZO-120. Besides,
$ZnCr_{2}O_{4}$ phase is identified in CZO-120. The transmittance
decreases again due to the scattering of photons by grain boundaries
and the secondary phase. On the other hand, $E_{U}$ is comparable
to the level of $Zn_{i}^{\bullet}/Zn_{i}^{\bullet\bullet}$, suggesting
that $Zn_{i}$ decides the Urbach tail. Hence, $Zn_{i}$ is the dominant
origin of carriers (\emph{n-type}) in the $ZnO$ films.

Due to increases in $E_{g}$ and/or grain boundaries, $\sigma_{dc}$
of ZnO films decreases after doping Cr. However, $\sigma_{dc}$ of
CZO-60 increases because of the highest concentrations of donors (Figure
\ref{fig:Photoluminescence-spectra-of}(d)). The transition of $\sigma_{dc}$
in 40\textminus 90°C is ascribed to the change of the electron origin.
Below 90°C, $Zn_{i}$ is the dominant carrier source, but with increasing
temperature, other defects are involved. Higher activation energies
of ZnO films increase after doping Cr, indicate stronger temperature-dependent
conduction barriers and thereby improved temperature stability. Therefore,
the method in this study is feasible to enhance temperature stability
with suitable amount of Cr doping and simultaneously maintain the
transmittance and DC conductivity at higher levels. In addition, $M\lyxmathsym{\textendash}H$
curves of Cr-doped ZnO films suggest diamagnetic behaviour influenced
by Cr concentration (see Figure S6), which needs to be further investigated.

\section{Conclusions}

Cr-doped ZnO films were fabricated using DC magnetron sputtering and
a new doping method of annealing two layers of metals in air is proposed.
ZnO grains preferentially grew along the \emph{c}-oriented (002) plane.
With increasing Cr content, a secondary phase of $ZnCr_{2}O_{4}$
was identified and the average grain size decreases from 56.34 nm
to 39.50 nm. Intrinsic point defects, that is, zinc interstitials,
oxygen vacancies, zinc vacancies, and oxygen interstitials, were detected
by PL spectra. The ratios of donors to zinc vacancies increase with
increasing Cr content, except for the highest amount. 

The band gap increases from 3.18 eV to 3.23 eV, but the transmittance
decreases from 91\% to 83\% (600 nm). The DC conductivity decreases
after doping Cr but considerably increases to a high level at an optimal
content. There are two different conduction processes in the ZnO films
and their activation energies increase after doping Cr, suggesting
enhanced temperature stability. In a word, the method of annealing
Cr-Zn layers is feasible to fabricate ZnO films with a passivation
layer and an optimal Cr doping level can help develop high-performance
transparent electrodes due to simultaneously maintaining higher transmittance
and conductivity. 

\subsubsection*{CRediT authorship contribution statement }

Men Guo: Conceptualization, visualization, data curation, investigation,
and writing -- original draft; Gilad Orr: Resources, editing, and
supervision; Paul Ben Ishai: Methodology, software, writing -- review
\& editing, supervision, funding acquisition, and project administration;
Xia Zhao: Funding acquisition; Shlomo Glasser: Resources. 

\subsubsection*{Declaration of Competing Interest}

The authors declare that they have no known competing financial interests
or personal relationships that could have influenced the work reported
in this paper.

\bibliographystyle{unsrt}
\bibliography{references}
\pagebreak{}

\part*{Supplemental Information}

\renewcommand{\figurename}{Figure S}\setcounter{figure}{0}

\begin{figure}[H]
\begin{centering}
\includegraphics[height=6cm]{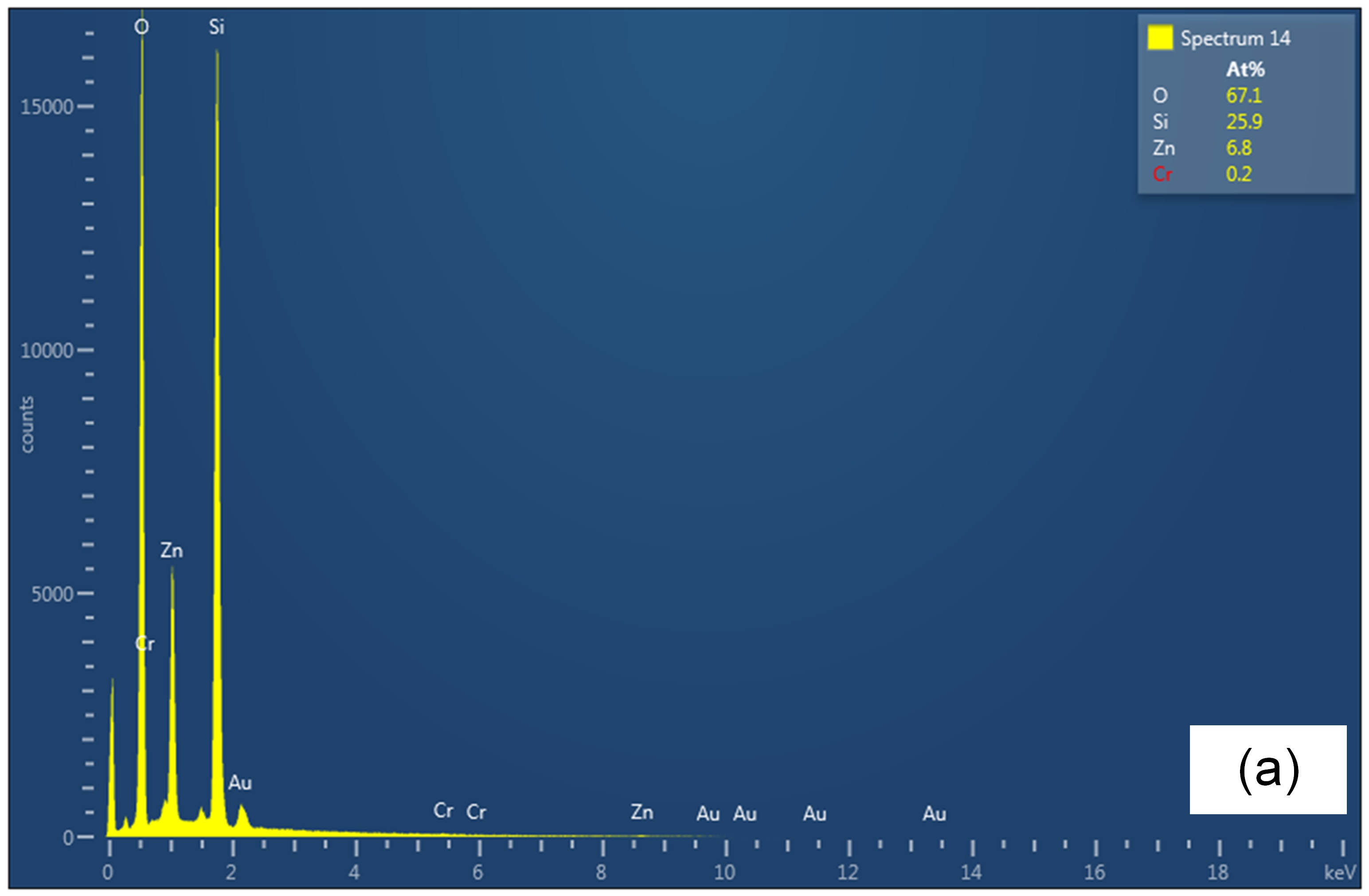}
\par\end{centering}
\begin{centering}
\includegraphics[height=6cm]{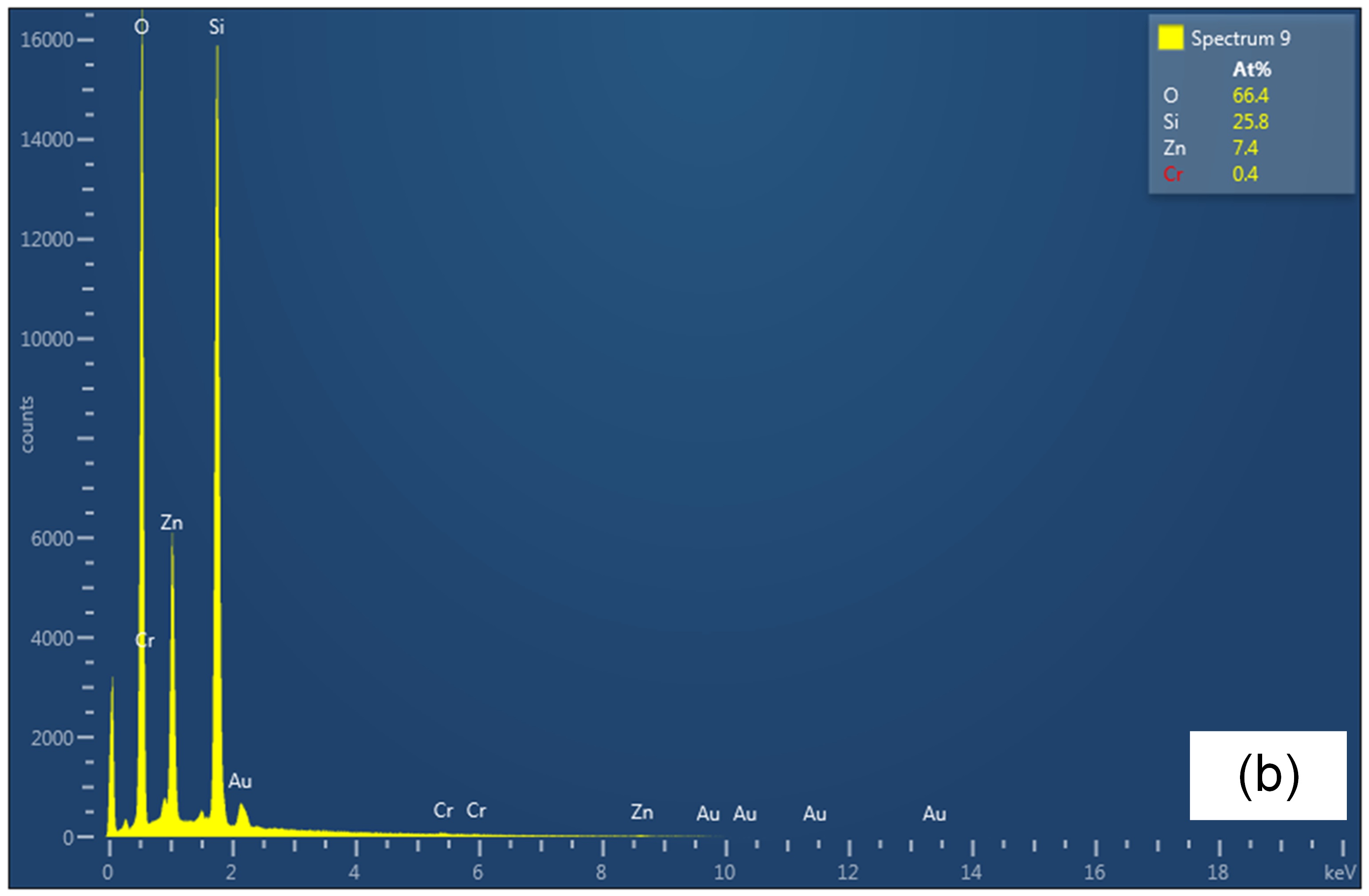}
\par\end{centering}
\caption{\label{fig:S1}EDS measurement results of Cr-doped ZnO films (a) CZO-60
(b) CZO-120}

\end{figure}

EDS measurement results of Cr-doped ZnO films are shown in Figure
S\ref{fig:S1}. Due to the ultrathin nature of the films (< 65 nm)
and low Cr doping levels, weak Cr signals (< 1 at\%) were detected
only in CZO-60 and CZO-120. No Cr signals were observed in CZO-0 (undoped)
or CZO-30 due to insufficient doping concentrations.
\pagebreak{}

\begin{figure}[H]
\begin{centering}
\includegraphics[height=4cm]{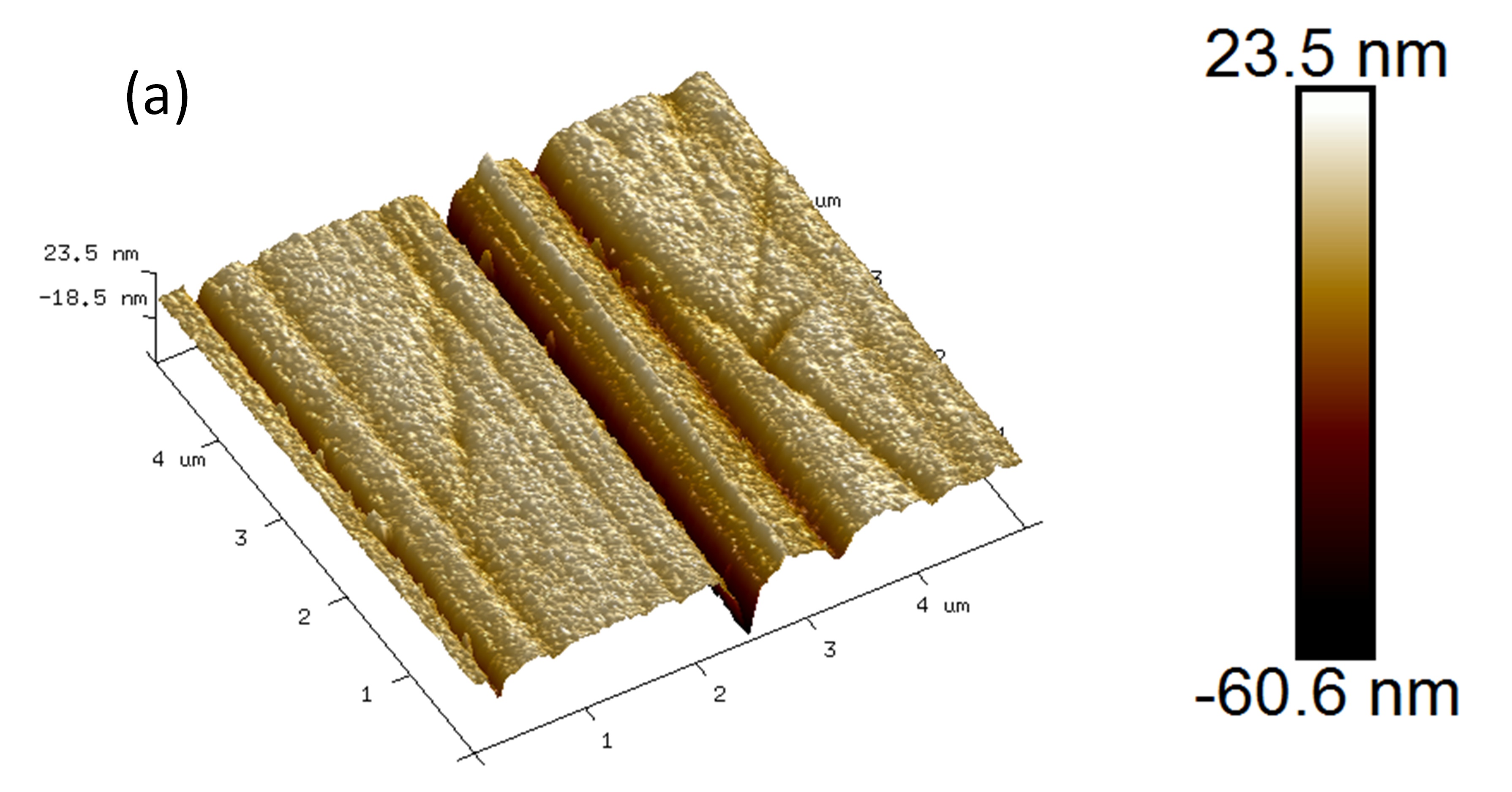} \includegraphics[height=4cm]{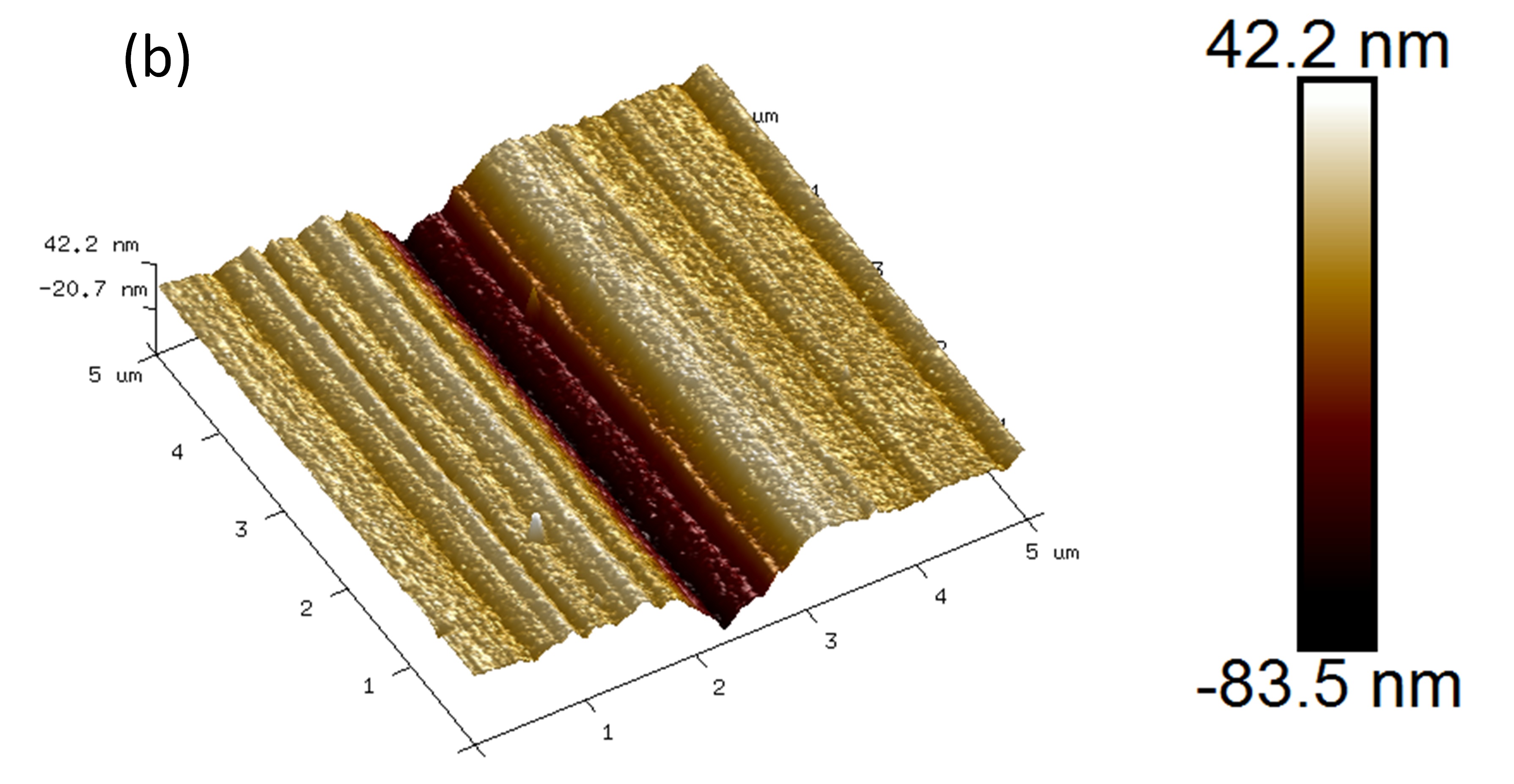}
\par\end{centering}
\begin{centering}
\includegraphics[height=4cm]{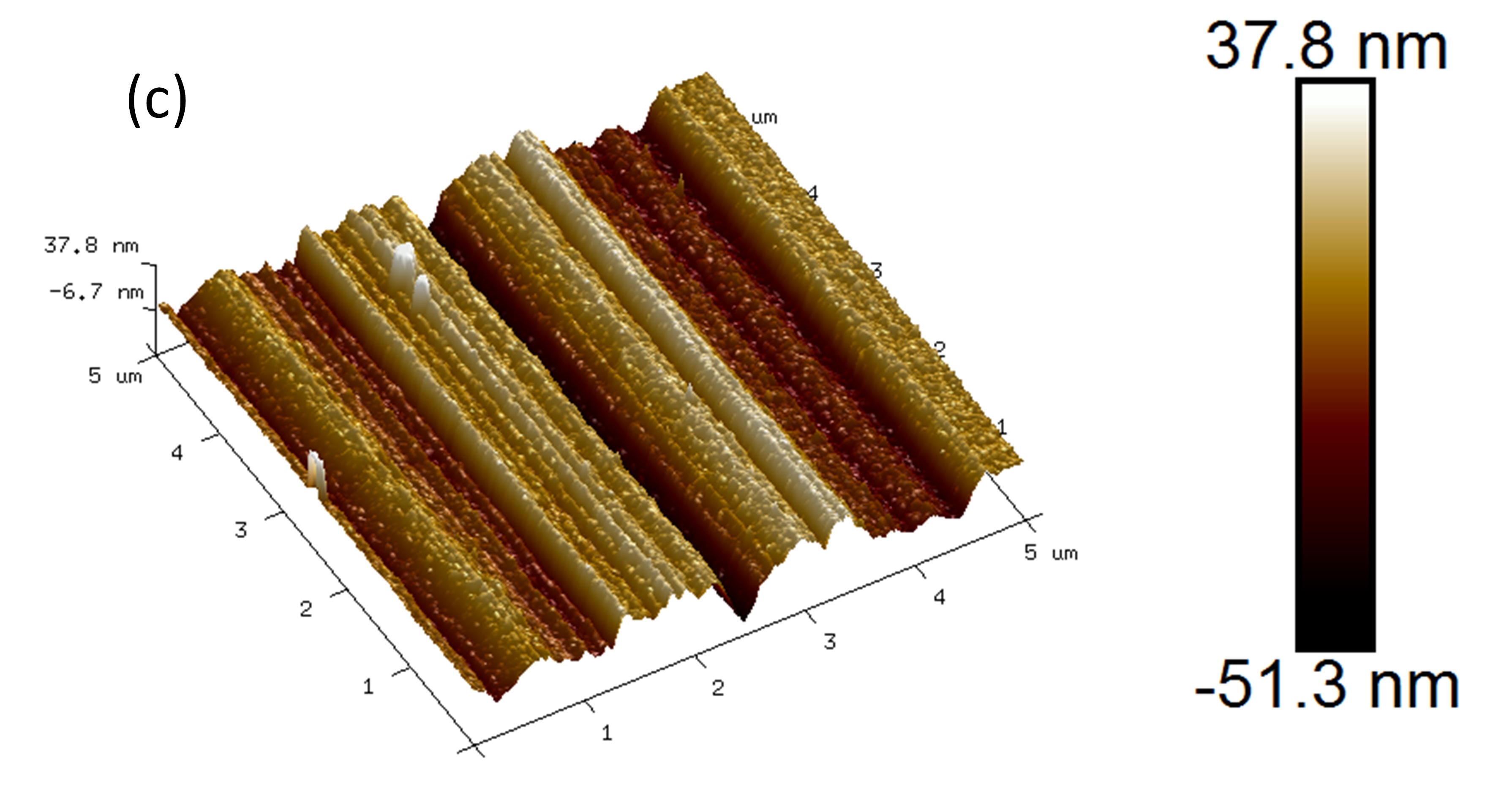} \includegraphics[height=4cm]{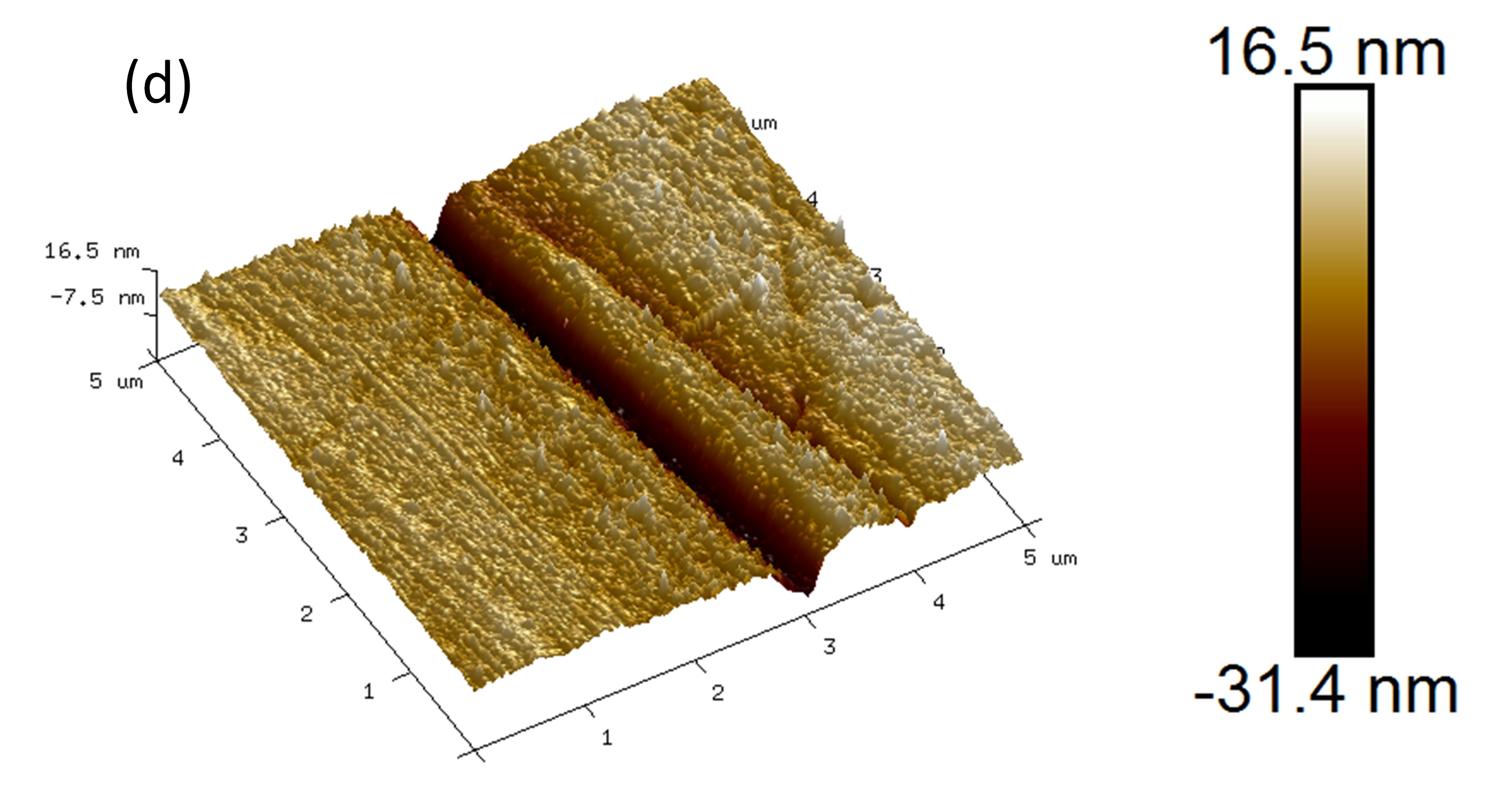}
\par\end{centering}
\begin{centering}
\includegraphics[scale=0.3]{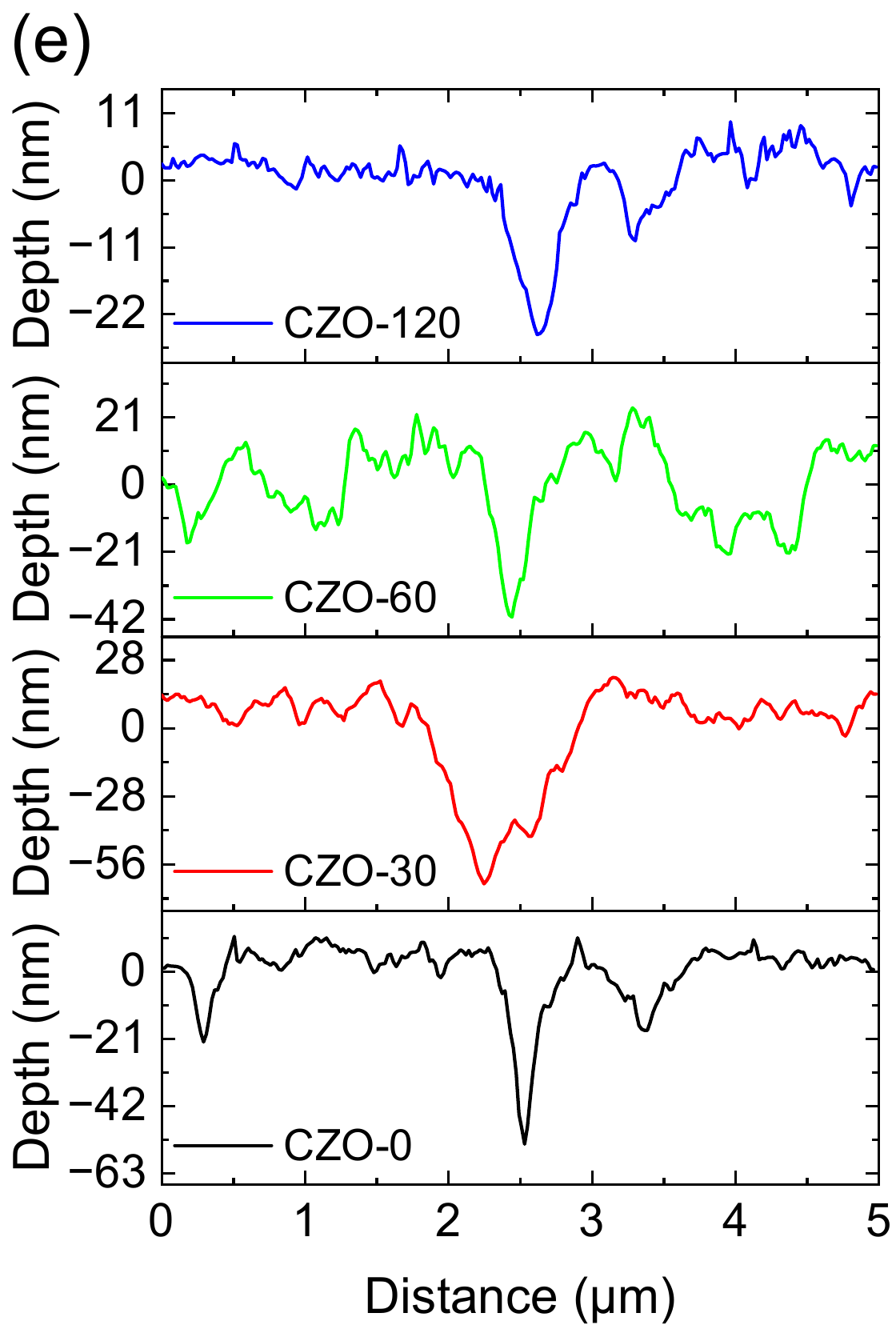}
\par\end{centering}
\caption{\label{fig:S2}AFM micrographs of Cr-doped ZnO films for thickness
measurements (a) CZO-0 (b) CZO-30 (c) CZO-60 (d) CZO-120 and (e) Thickness
profiles.}
\end{figure}

Film thicknesses (\emph{t}), measured via atomic force microscopy
(AFM) by scratching films and analysing depth profiles, are shown
in Figure S\ref{fig:S2}. \emph{t} is 53.8, 63.9, 41.3, and 25.3 nm
for CZO-0, CZO- 30, CZO-60, and CZO-120, respectively.

\pagebreak{}

\begin{figure}[H]
\begin{centering}
\includegraphics[height=5.5cm]{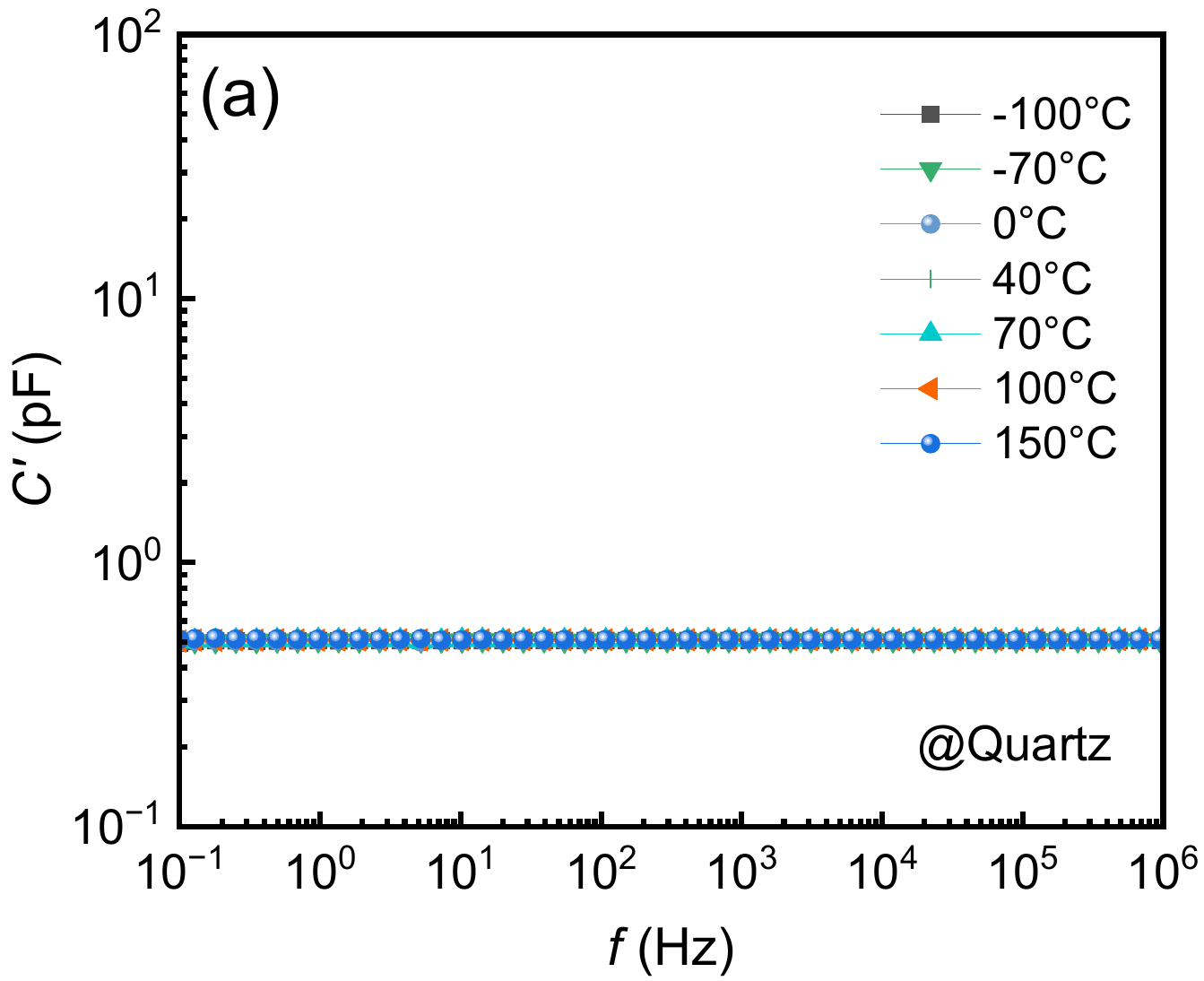} \includegraphics[height=5.5cm]{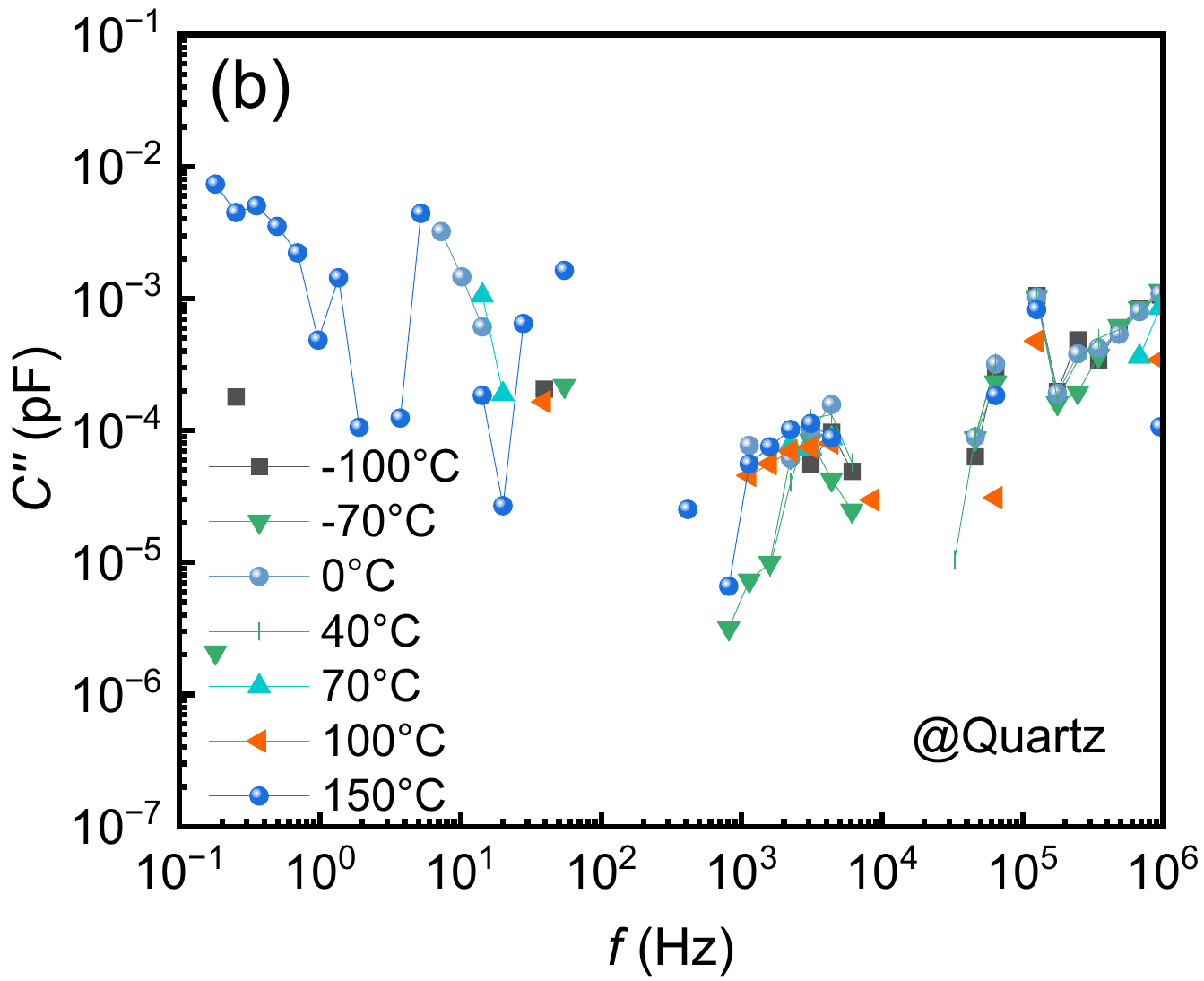}
\par\end{centering}
\caption{\label{fig:S3}Dielectric spectra of the quartz substrate (a) the
real part $C'$ (b) the imaginary part $C''$.}

\end{figure}

Dielectric spectra of the quartz substrate are shown in Figure S\ref{fig:S3}.
$C'$ does not show temperature and frequency dependence and the value
is approximately 0.51 pF. $C"$ is considerably low ( <~$10^{-2}\,pF$)
and at some frequencies cannot be precisely measured.

\pagebreak{}

\begin{figure}[H]
\begin{centering}
\includegraphics[height=5cm]{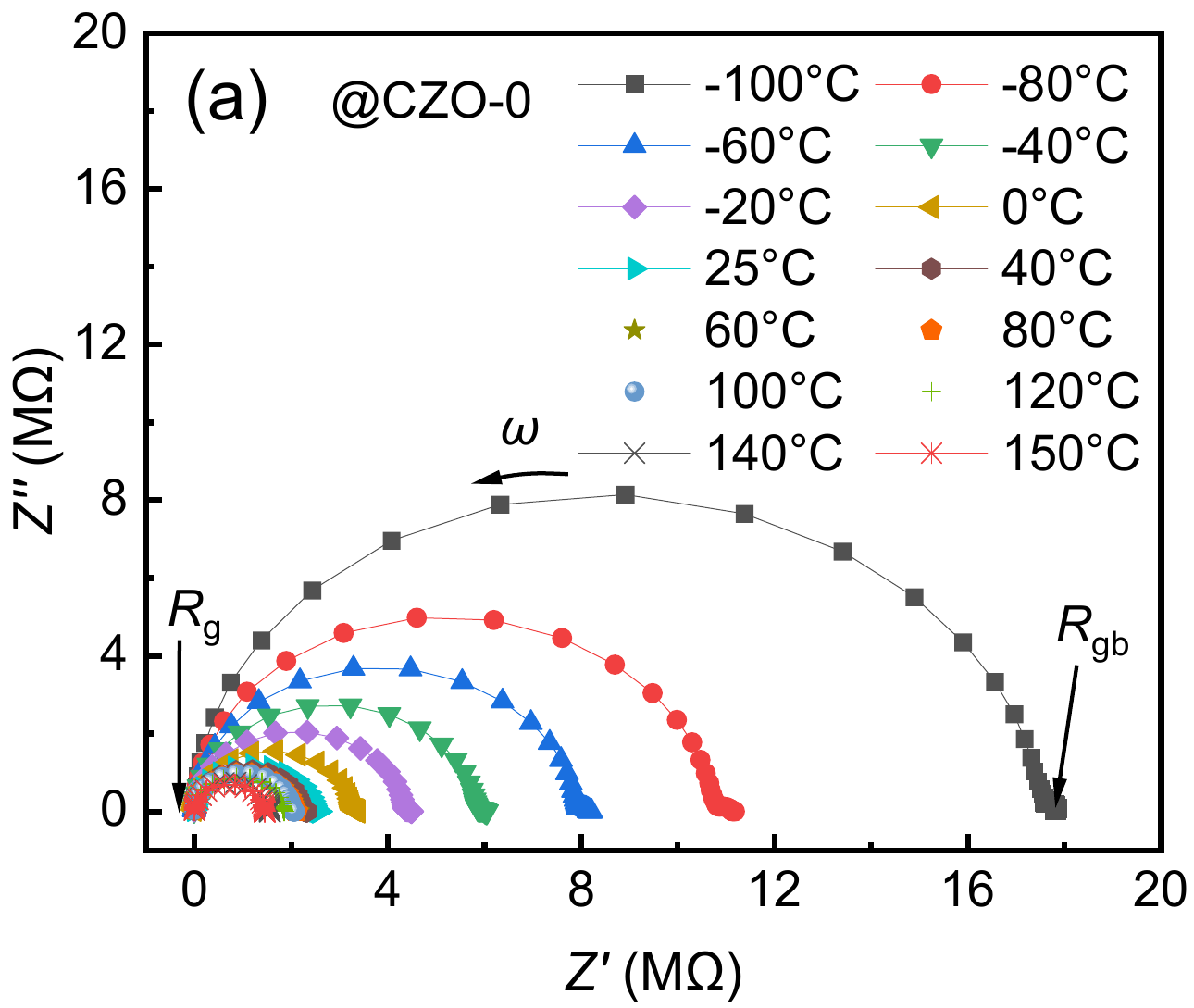} \includegraphics[height=5cm]{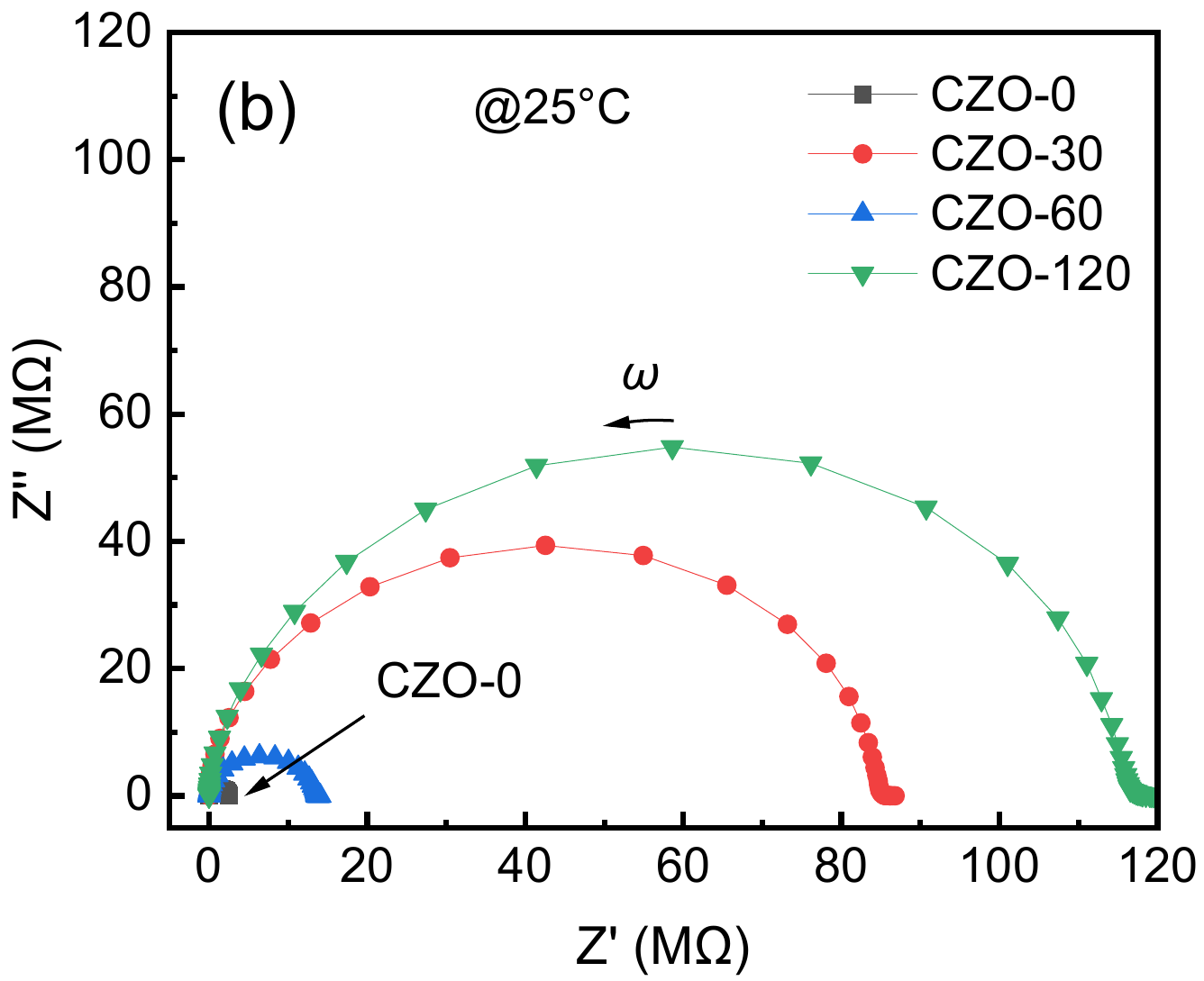}
\par\end{centering}
\begin{centering}
\includegraphics[height=5cc]{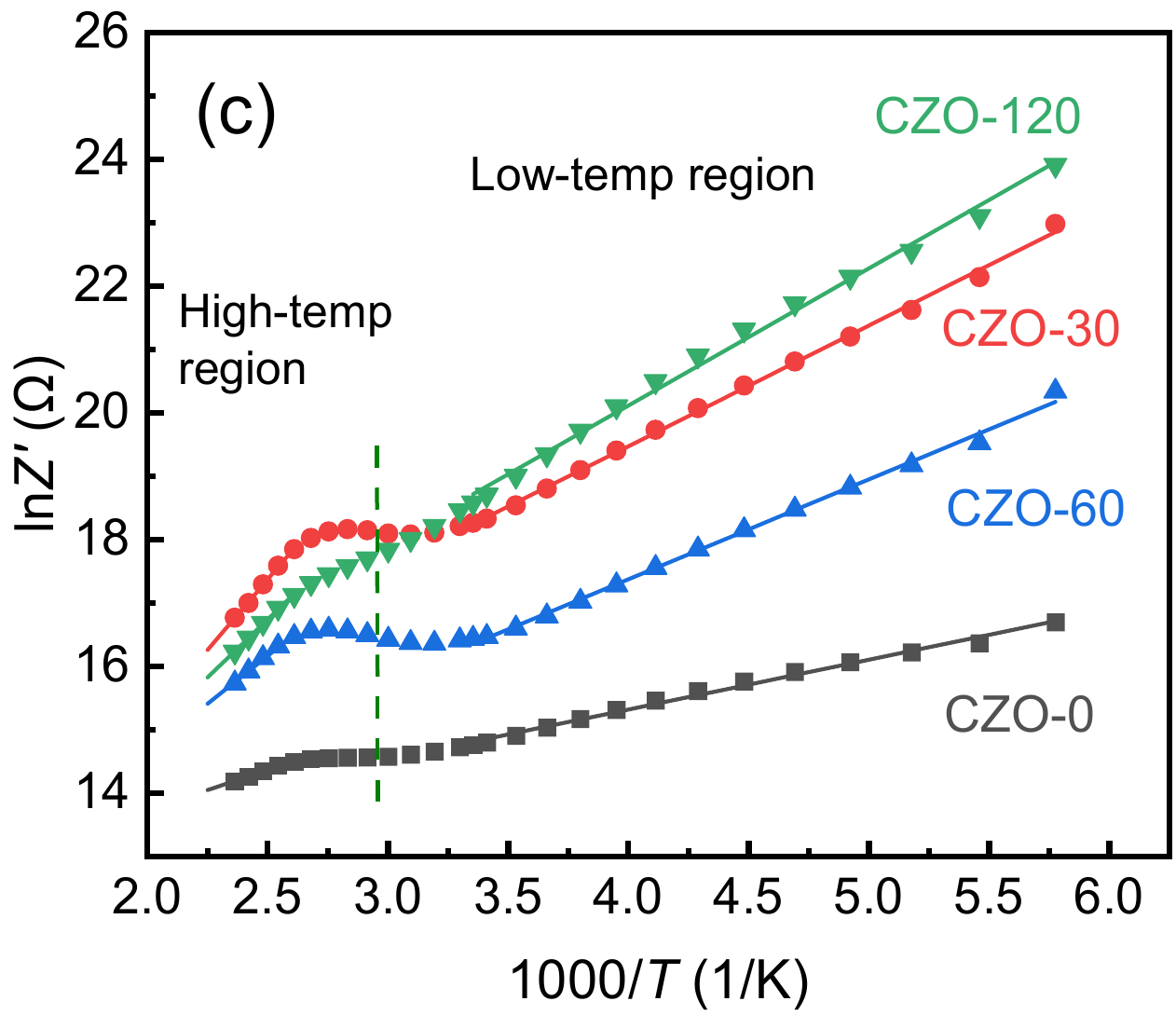}
\par\end{centering}
\caption{\label{fig:S4}The Nyquist plots of (a) CZO-0 in \textminus 100\textminus 150°C
(b) all the films at 25°C, and (c) the Arrhenius plot.}

\end{figure}

The Nyquist plots of Cr-doped ZnO films are shown in Figure S\ref{fig:S4}.
The impedance ($Z^{*}$) of ZnO films can be expressed as

\begin{align}
Z^{*} & =Z'-iZ''=R_{g}+\left(\frac{1}{R_{gb}}+i\omega C_{gb}\right)^{-1}\\
 & =R_{g}+\frac{R_{gb}}{1+(\omega R_{gb}C_{gb})^{2}}-iR_{gb}\frac{\omega R_{gb}C_{gb}}{1+(\omega R_{gb}C_{gb})^{2}}\nonumber 
\end{align}

where $Z'$ and $Z''$ are the real part and the imaginary part of
$Z^{*}$ respectively, $R_{g}$ is the resistance of ZnO grains, and
$R_{gb}$ and $C_{gb}$ are the resistance and the capacitance of
ZnO grain boundaries, respectively. If $\omega$ tends to 0, $Z''$
approximately equals to $R_{g}+R_{gb}$. If $\omega$ tends to $\infty$,
$Z'$ approximately equals to $R_{g}$. As $R_{gb}\gg R_{g}$, the
right intercept on the $Z'$ axis can be used to reflect $R_{gb}$.
With increasing temperature (\textminus 100\textminus 150°C), $R_{gb}$
of CZO-0 dramatically decreases from 18 M\textgreek{W} to less than
2 M\textgreek{W}, as shown in Figure S\ref{fig:S4}(a). With increasing
Cr content, $R_{gb}$ of ZnO films increase from 2.58 M\textgreek{W}
to 118 M\textgreek{W}, shown in Figure S\ref{fig:S4}(b). The Arrhenius
plot is present in Figure S\ref{fig:S4}(c). There are also two linear
regions at low temperatures (< 40°C) and high temperatures (> 90°C),
as reflected in the DC conduction.
\pagebreak{}

\begin{figure}[H]
\begin{centering}
\includegraphics[height=6cm]{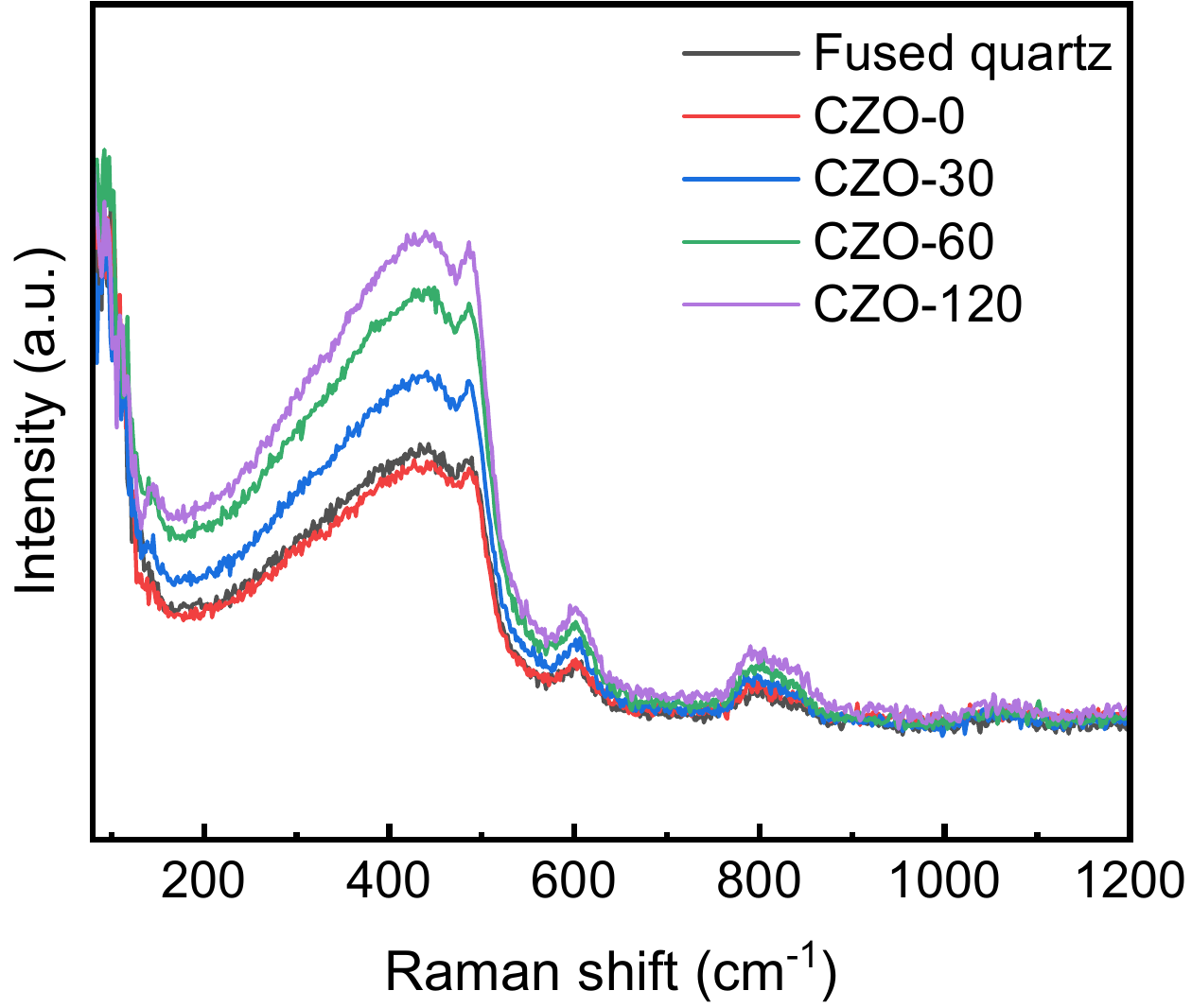}
\par\end{centering}
\caption{\label{fig:S5}Raman spectra of the Cr-doped ZnO films.}

\end{figure}

Raman spectra of the Cr-doped ZnO films are shown in Figure S\ref{fig:S5}.
Because the films are ultrathin, no Raman signals from the films were
detected. The measured spectra resemble those of the fused quartz
substrates.

\pagebreak{}

\begin{figure}[H]
\begin{centering}
\includegraphics[height=5cm]{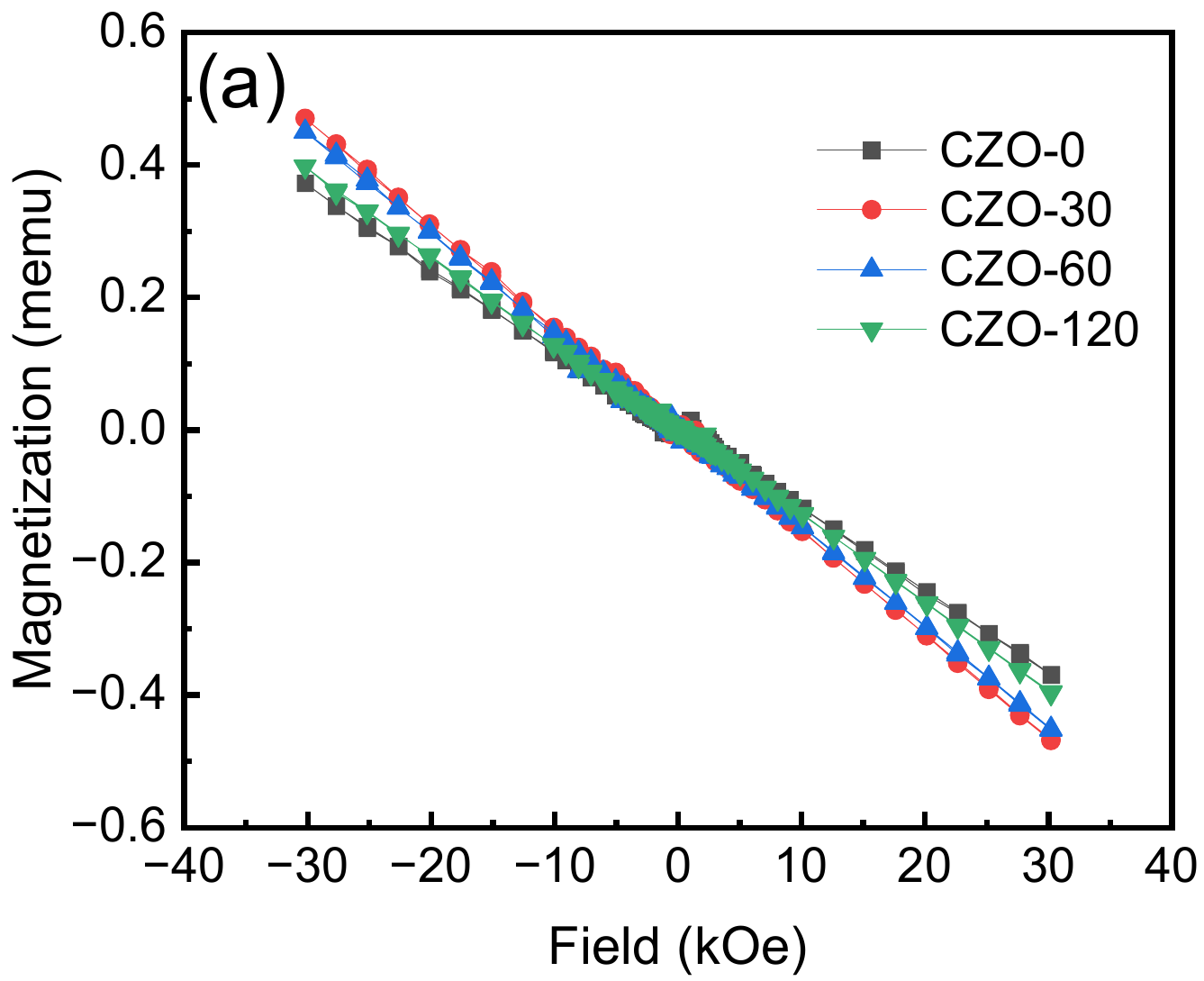} \includegraphics[height=5cm]{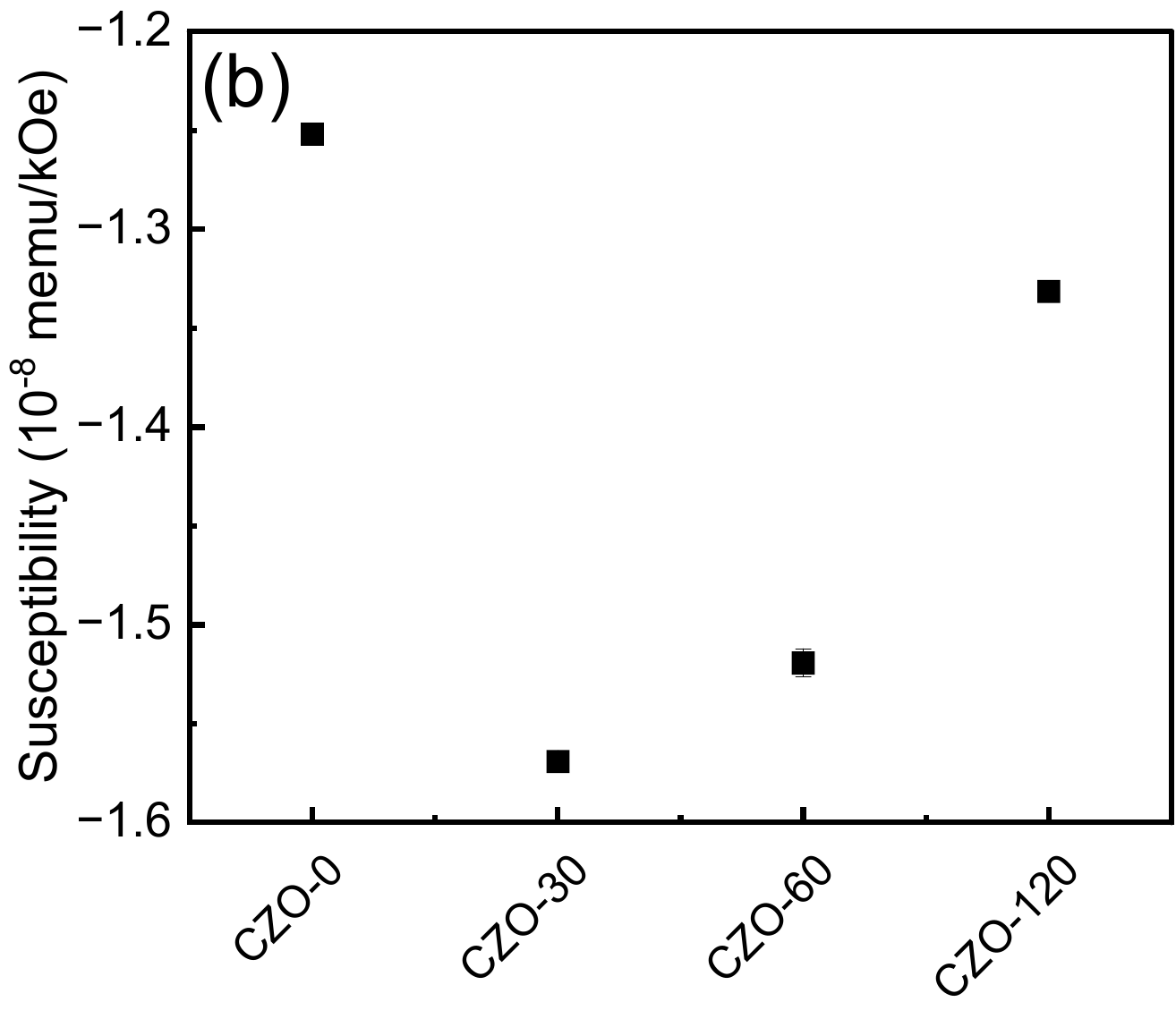}
\par\end{centering}
\caption{\label{fig:S6}(a) $M\lyxmathsym{\textendash}H$ curves and (b) magnetic
susceptibility of Cr-doped ZnO films at room temperature}

\end{figure}

Magnetization--magnetic field ($M\lyxmathsym{\textendash}H$) curves
of Cr-doped ZnO films are shown in Figure S\ref{fig:S6}(a). The films
do not show ferromagnetic behaviour but negative susceptibility, as
shown in Figure S\ref{fig:S6}(b), suggesting diamagnetic properties.
The results could be caused either by low Cr contents or diamagnetic
contributions of the fused quartz substrate.

\end{document}